%% file: main.tex
\documentclass[sigconf]{acmart}
\AtBeginDocument{%
  \providecommand\BibTeX{{%
    \normalfont B\kern-0.5em{\scshape i\kern-0.25em b}\kern-0.8em\TeX}}}

\newcommand{\Model}{\texttt{DaL}} 

\usepackage[ruled, linesnumbered, vlined, commentsnumbered]{algorithm2e}
\usepackage{xcolor,colortbl}
\usepackage{multirow}
\usepackage{subcaption} 
\usepackage{tcolorbox}
\usepackage{csquotes}
\usepackage{adjustbox}
\usepackage{balance}
\definecolor{beaublue}{rgb}{0.74, 0.83, 0.9}
\newtcolorbox{quotebox}{colback=beaublue,boxrule=0.4pt,colframe=black,fonttitle=\bfseries,top=2pt,bottom=2pt}

\newcolumntype{P}[1]{>{\centering\arraybackslash}p{#1}}

\definecolor{one}{HTML}{2b7bba}
\definecolor{two}{HTML}{d52221}

 \newcommand{\markone}[4]{\begin{adjustbox}{max width=.1\textwidth}\begin{picture}(20,5)
    {\linethickness{0.2mm}\color{one}\put(#1,3){\line(1,0){#2}}\color{one}\put(#3,3){\circle*{4}}}\end{picture}\end{adjustbox}}

    \newcommand{\marktwo}[4]{\begin{adjustbox}{max width=.1\textwidth}\begin{picture}(20,5)
    {\linethickness{0.2mm}\color{two}\put(#1,3){\line(1,0){#2}}\color{two}\put(#3,0.5){\large$\star$}}\end{picture}\end{adjustbox}}

\DeclareMathAlphabet\mathbfcal{OMS}{cmsy}{b}{n}
\SetKwInput{kwDeclare}{Declare}

\setcopyright{rightsretained}
\acmPrice{}
\acmDOI{10.1145/3611643.3616334}
\acmYear{2023}
\copyrightyear{2023}
\acmSubmissionID{fse23main-p955-p}
\acmISBN{979-8-4007-0327-0/23/12}
\acmConference[ESEC/FSE '23]{Proceedings of the 31st ACM Joint European Software Engineering Conference and Symposium on the Foundations of Software Engineering}{December 3--9, 2023}{San Francisco, CA, USA}
\acmBooktitle{Proceedings of the 31st ACM Joint European Software Engineering Conference and Symposium on the Foundations of Software Engineering (ESEC/FSE '23), December 3--9, 2023, San Francisco, CA, USA}
\received{2023-02-02}
\received[accepted]{2023-07-27}

\begin{document}

\title{Predicting Software Performance with Divide-and-Learn}

\author{Jingzhi Gong}
\affiliation{%
  \institution{Loughborough University}
  \city{Leicestershire\\}
  \country{United Kingdom}
 }
\email{j.gong@lboro.ac.uk}

\author{Tao Chen}
\authornote{Tao Chen is the corresponding author}
\affiliation{
  \institution{University of Birmingham}
  \city{Birmingham\\}
  \country{United Kingdom}
}
\email{t.chen@bham.ac.uk}

\input{abstract}




\maketitle

\input{introduction}

\input{background}

\input{framework}
\input{study}

\input{evaluation}

\input{discussion}

\input{related}
\input{conclusion}

\balance
\bibliographystyle{ACM-Reference-Format}
\bibliography{references}

\end{document}

%% file: abstract.tex
\begin{abstract}

Predicting the performance of highly configurable software systems is the foundation for performance testing and quality assurance. To that end, recent work has been relying on machine/deep learning to model software performance. However, a crucial yet unaddressed challenge is how to cater for the sparsity inherited from the configuration landscape: the influence of configuration options (features) and the distribution of data samples are highly sparse. 

In this paper, we propose an approach based on the concept of ``divide-and-learn'', dubbed \Model. The basic idea is that, to handle sample sparsity, we divide the samples from the configuration landscape into distant divisions, for each of which we build a regularized Deep Neural Network as the local model to deal with the feature sparsity. A newly given configuration would then be assigned to the right model of division for the final prediction. 

Experiment results from eight real-world systems and five sets of training data reveal that, compared with the state-of-the-art approaches, \Model~performs no worse than the best counterpart on 33 out of 40 cases (within which 26 cases are significantly better) with up to $1.94\times$ improvement on accuracy; requires fewer samples to reach the same/better accuracy; and producing acceptable training overhead. Practically, \Model~also considerably improves different global models when using them as the underlying local models, which further strengthens its flexibility. To promote open science, all the data, code, and supplementary figures of this work can be accessed at our repository: \texttt{\textcolor{blue}{\url{https://github.com/ideas-labo/DaL}}}.

\end{abstract}

%% file: introduction.tex
\section{Introduction}
\label{sec:introduction}

\begin{displayquote}
\textit{``What will be the implication on runtime if we deploy that configuration?''} 
\end{displayquote}

The above is a question we often hear from our industrial partners. Indeed, software performance, such as latency, runtime, and energy consumption, is one of the most critical concerns of software systems that come with a daunting number of configuration options, e.g., \textsc{x264} (a video encoder) allows one to adjust 16 options to influence its runtime. To satisfy the performance requirements, it is essential for software engineers to understand what performance can be obtained under a given configuration before the deployment. This not only enables better decisions on configuration tuning~\cite{10.1145/3571853} but also reduces the efforts of configuration testing~\cite{DBLP:conf/wosp/FermeP17}.

To achieve the above, one way is to directly profile the software system for all possible configurations when needed. This, however, is impractical, because (1) the number of possible configurations may be too high~\cite{DBLP:conf/mascots/JamshidiC16,DBLP:conf/sigsoft/0001Chen21,DBLP:conf/sigsoft/SiegmundGAK15, DBLP:conf/icse/HaZ19}. For example, \textsc{HIPA$^{cc}$} (a compiler for image processing) has more than 10,000 possible configurations. (2) Even when such a number is small, the profiling of a single configuration can still be rather expensive~\cite{DBLP:conf/mascots/JamshidiC16,DBLP:conf/sigsoft/0001Chen21}: Wang \textit{et al.}~\cite{DBLP:conf/sigsoft/WangHJK13} report that it could take weeks of running time to benchmark and profile even a simple system. Therefore, an accurate performance model that can predict the expected performance of a newly given configuration is of high demand. 

With the increasing complexity of modern software, the number of configurable options continues to expand and the interactions between options become more complicated, leading to significant difficulty in predicting the performance accurately~\cite{DBLP:conf/icse/SiegmundKKABRS12,DBLP:journals/tse/ChenB17}. Recently, machine learning models have been becoming the promising method for this regression problem as they are capable of modeling the complex interplay between a large number of variables by observing patterns from data~\cite{DBLP:conf/icse/WeberAS21,DBLP:conf/icse/VelezJSAK21, DBLP:conf/wosp/HanYP21,DBLP:conf/sigsoft/SiegmundGAK15,DBLP:conf/esem/ShuS0X20,DBLP:journals/sqj/SiegmundRKKAS12,DBLP:conf/icse/HaZ19}.

However, since machine learning modeling is data-driven, the characteristics and properties of the measured data for configurable software systems pose non-trivial challenges to the learning, primarily because it is known that the configuration landscapes of the systems do not follow a ``smooth'' shape~\cite{DBLP:conf/mascots/JamshidiC16}. For example, adjusting between different cache strategies can drastically influence the performance, but they are often represented as a single-digit change on the landscape~\cite{DBLP:journals/corr/abs-1801-02175,DBLP:conf/sigsoft/0001Chen21,DBLP:conf/seams/Chen22}. This leads to the notion of sparsity in two aspects:

\begin{itemize}
    \item Only a small number of configuration options can significantly influence the performance, hence there is a clear \textbf{\textit{feature sparsity}} involved~\cite{DBLP:conf/nips/HuangJYCMN10, DBLP:conf/sigsoft/SiegmundGAK15, DBLP:conf/icse/HaZ19, DBLP:conf/icse/VelezJSAK21}.
    \item The samples from the configuration landscape tend to form different divisions with diverse values of performance and configuration options, especially when the training data is limited due to expensive measurement---a typical case of \textbf{\textit{sample sparsity}}~\cite{DBLP:journals/corr/abs-2202-03354,DBLP:conf/icml/LiuCH20,DBLP:conf/icml/ShibagakiKHT16}. This is particularly true when not all configurations are valid~\cite{DBLP:conf/sigsoft/SiegmundGAK15}.
\end{itemize}

Existing work has been primarily focusing on addressing feature sparsity, through using tree-liked model~\cite{DBLP:conf/kbse/GuoCASW13}; via feature selection~\cite{DBLP:journals/fgcs/LiLTWHQD19, DBLP:conf/mascots/GrohmannEEKKM19,DBLP:journals/tse/ChenB17}; or deep learning~\cite{DBLP:conf/icse/HaZ19, DBLP:journals/jmlr/GlorotBB11,DBLP:conf/esem/ShuS0X20}. However, the sample sparsity has almost been ignored, which can still be a major obstacle to the effectiveness of machine learning-based performance model.

To address the above gap, in this paper, we propose \Model, an approach to model software performance via the concept of ``divide-and-learn''. The basic idea is that, to handle sample sparsity, we divide the samples (configurations and their performance) into different divisions, each of which is learned by a local model. In this way, the highly sparse samples can be split into different locally smooth regions of data samples, and hence their patterns and feature sparsity can be better captured.

In a nutshell, our main contributions are:
\begin{enumerate}

    \item We formulate, on top of the regression of performance, a new classification problem without explicit labels.
    \item We extend Classification and Regression Tree (CART)~\cite{loh2011classification} as a clustering algorithm to ``divide'' the samples into different divisions with similar characteristics, for each of which we build a local regularized Deep Neural Network (rDNN)~\cite{goodfellow2016deep}.
    \item Newly given configurations would be assigned into a division inferred by a Random Forest classifier~\cite{DBLP:conf/icdar/Ho95}, which is trained using the pseudo labeled data from the CART. The rDNN model of the assigned division would be used for the final prediction thereafter.
    \item Under eight systems with diverse performance attributes, scale, and domains, as well as five different training sizes, we evaluate \Model~against four state-of-the-art approaches and with different underlying local models.

\end{enumerate}

The experiment results are encouraging: compared with the best state-of-the-art approach, we demonstrate that \Model

\begin{itemize}
        \item achieves no worse accuracy on 33 out of 40 cases with 26 of them being significantly better. The improvements can be up to $1.94\times$ against the best counterpart; 
    \item uses fewer samples to reach the same/better accuracy.
    \item incurs acceptable training time considering the improvements in accuracy. 
\end{itemize}

Interestingly, we also reveal that:

\begin{itemize}
    \item \Model~can considerably improve the accuracy of an arbitrarily given model when it serves as the local model for each division compared with using the model alone as a global model (which is used to learn the entire training dataset). However, the original \Model~with rDNN as the local model still produces the most accurate results.
    \item \Model's error tends to correlate quadratically with its only parameter $d$ that sets the number of divisions. Therefore, a middle value (between 0 and the bound set by CART) can reach a good balance between handling sample sparsity and providing sufficient training data for the local models, e.g., $d=1$ or $d=2$ (2 or 4 divisions) in this work.

\end{itemize}

This paper is organized as follows: Section~\ref{sec:background} introduces the problem formulation and the notions of sparsity in software performance learning. Section~\ref{sec:framework} delineates the tailored problem formulation and our detailed designs of \Model. Section~\ref{sec:setup} presents the research questions and the experiment design, followed by the analysis of results in Section~\ref{sec:evaluation}. The reasons why \Model~works, its strengths, limitations, and threats to validity are discussed in Section~\ref{sec:discussion}. Section~\ref{sec:related},~\ref{sec:conclusion}, and~\ref{sec:data} present the related work, conclude the paper, and elaborate data availability, respectively.

%% file: background.tex
\section{Background and Motivation}
\label{sec:background}

In this section, we introduce the background and the key observations that motivate this work.


\subsection{Problem Formulation}


In the software engineering community, the question introduced at the beginning of this paper has been most commonly addressed by using various machine learning models (or at least partially)~\cite{DBLP:journals/tse/YuWHH19,DBLP:conf/splc/TempleAPBJR19,DBLP:journals/tosem/ChenLBY18,DBLP:journals/corr/abs-1801-02175,DBLP:conf/esem/HanY16}, in a data-driven manner that relies on observing the software’s actual behaviors and builds a statistical model to predict the performance without heavy human intervention~\cite{DBLP:books/daglib/0020252}. 


Formally, modeling the performance of software with $n$ configuration options is a regression problem that builds:
\begin{equation}
    \mathcal{P} = f(\mathbfcal{S})\text{, } \mathcal{P}\in\mathbb{R}
    \label{eq:prob}
\end{equation}
whereby $\mathbfcal{S}$ denotes the training samples of configuration-performance pairs, such that $\mathbf{\overline{x}} \in \mathbfcal{S}$. $\mathbf{\overline{x}}$ is a configuration and $\mathbf{\overline{x}}=(x_{1},x_{2},\cdots,x_{n})$, where each configuration option $x_{i}$ is either binary or categorical/numerical. The corresponding performance is denoted as $\mathcal{P}$. 


The goal of machine learning-based modeling is to learn a regression function $f$ using all training data samples such that for newly given configurations, the predicted performance is as close to the actual performance as possible. 

\input{Figures/3d_scatter}

\subsection{Sparsity in Software Performance Learning}
\label{sec:motivation}

It has been known that the configuration space for software systems is generally rugged and sparse with respect to the configuration options~\cite{DBLP:conf/mascots/JamshidiC16,DBLP:conf/sigsoft/0001Chen21,DBLP:conf/icse/HaZ19,DBLP:conf/wcre/Chen22} --- \textbf{\textit{feature sparsity}}, which refers to the fact that only a small number of configuration options are prominent to the performance. We discover that, even with the key options that are the most influential to the performance, the samples still do not exhibit a ``smooth'' distribution over the configuration landscape. Instead, they tend to be spread sparsely: those with similar characteristics can form arbitrarily different divisions, which tend to be rather distant from each other. This is a typical case of high \textbf{\textit{sample sparsity}}~\cite{DBLP:journals/corr/abs-2202-03354,DBLP:conf/icml/LiuCH20,DBLP:conf/icml/ShibagakiKHT16} and it is often ignored in existing work for software performance learning.

In Figure~\ref{fig:3d-exp}, we show examples of the configuration samples measured from four real-world software systems. Clearly, we see that they all exhibit a consistent pattern\footnote{Similar pattern has been registered on all systems studied in this work.}---the samples tend to form different divisions with two properties:

\begin{itemize}
    \item \textbf{Property 1:} configurations in the same division share closer performance values with smoother changes but those in-between divisions exhibit drastically different performance and can change more sharply.
    \item \textbf{Property 2:} configurations in the same division can have a closer value on at least one key option than those from the different divisions.
\end{itemize}

In this regard, the values of performance and key configuration options determine the characteristics of samples. In general, such a high sample sparsity is caused by two reasons: (1) the inherited consequence of high feature sparsity and (2) the fact that not all configurations are valid because of the constraints (e.g., an option can be used only if another option has been turned on)~\cite{DBLP:conf/sigsoft/SiegmundGAK15}, thereby there are many ``empty areas'' in the configuration landscape.

When using machine learning models to learn concepts from the above configuration data, the model needs to (1) handle the complex interactions between the configuration options with high feature sparsity while (2) capture the diverse characteristics of configuration samples over all divisions caused by the high sample sparsity, e.g., in Figure~\ref{fig:3d-exp}, where samples in different divisions have diverged performance ranges. For the former challenge, there have been some approaches proposed to target such, such as \texttt{DeepPerf}~\cite{DBLP:conf/icse/HaZ19} and \texttt{Perf-AL}~\cite{DBLP:conf/esem/ShuS0X20}. However, very little work has aimed to address the latter which can be the main obstacle for a model to learn and generalize the data for predicting the performance of the newly-given configuration. This is because those highly sparse samples increase the risk for models to overfit the training data, for instance by memorizing and biasing values in certain respective divisions~\cite{DBLP:journals/corr/abs-2202-03354}, especially considering that we can often have limited samples from the configuration landscape due to the expensive measurement of configurable systems. The above is the main motivation of this work, for which we ask: how can we improve the accuracy of predicting software performance under such a high sample sparsity?

%% file: Figures/3d_scatter.tex
\begin{figure}[!t]
\centering
\footnotesize

\begin{subfigure}{.49\columnwidth}
  \centering
  \includegraphics[width=\linewidth]{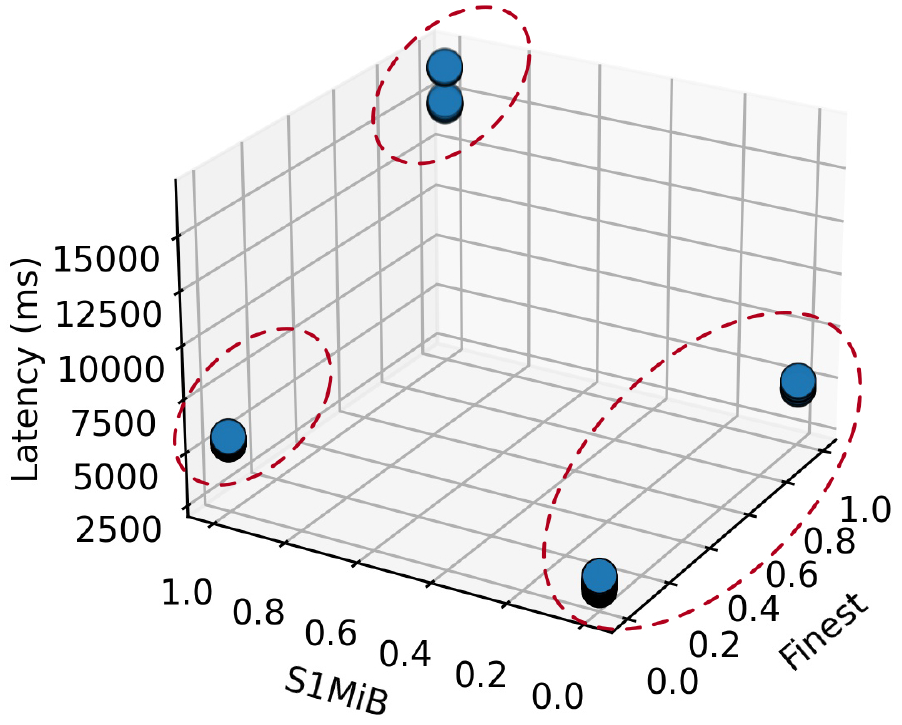} 
  \caption{\textsc{BDB-J}}
  \label{fig:depth-Apache}
\end{subfigure}
\begin{subfigure}{.49\columnwidth}
  \centering
  \includegraphics[width=\linewidth]{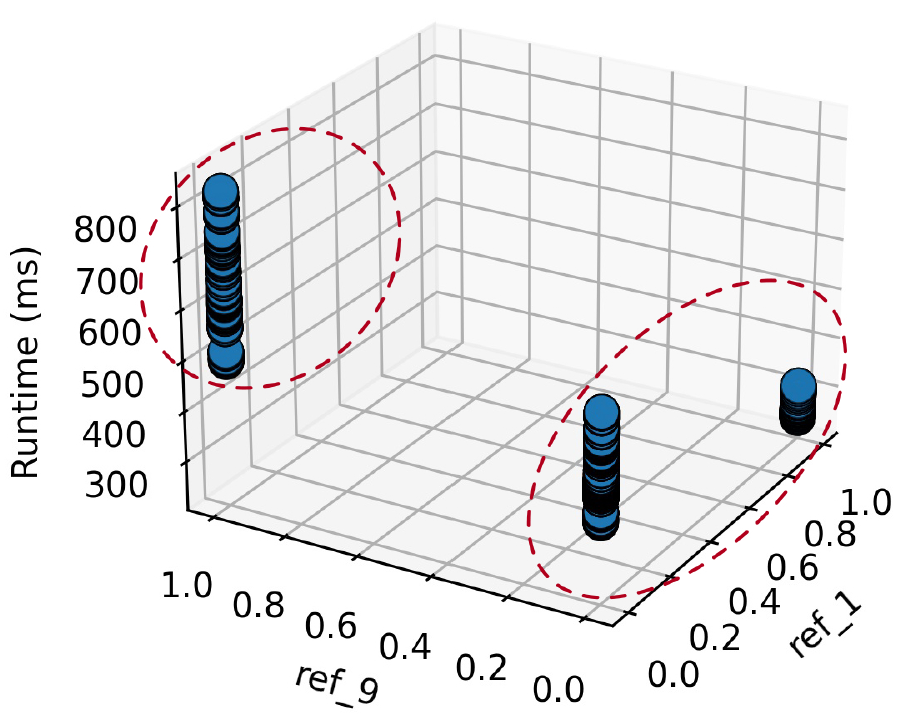} 
  \caption{\textsc{x264}}
  \label{fig:depth-BDBC}
\end{subfigure}

\begin{subfigure}{.49\columnwidth}
  \centering
  \includegraphics[width=\linewidth]{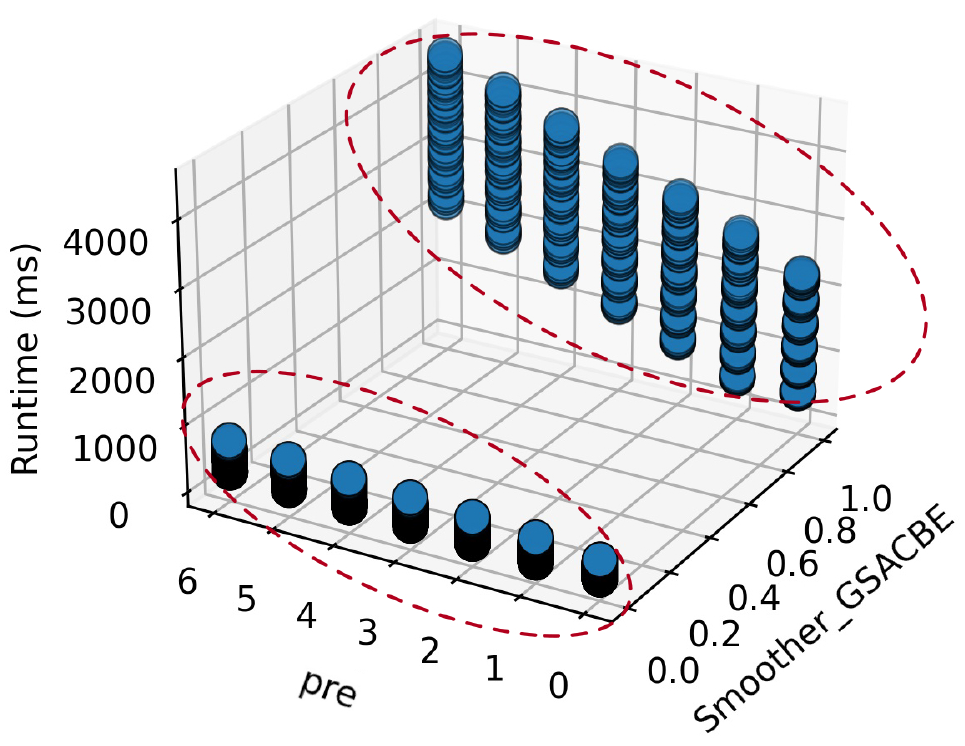} 
  \caption{\textsc{HSMGP}}
  \label{fig:depth-BDBJ}
\end{subfigure}
\begin{subfigure}{.49\columnwidth}
  \centering
  \includegraphics[width=\linewidth]{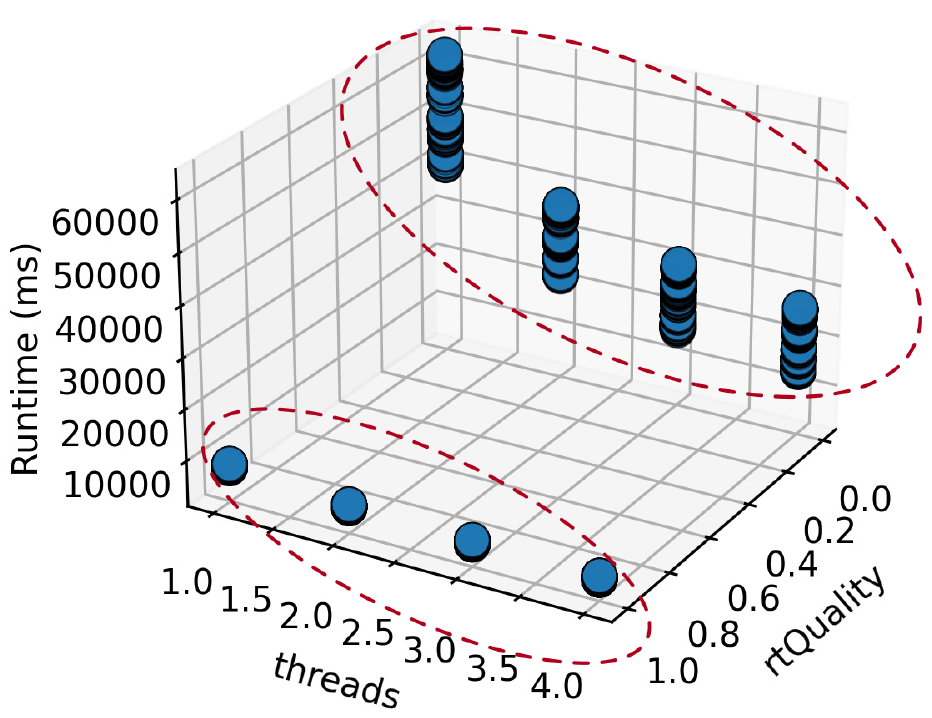}  
  \caption{\textsc{VP8}}
  \label{fig:depth-x264}
\end{subfigure}
  \caption{Projection of configurations in the landscape with respect to the performance and two most important options (the divisions are circled).}
     \label{fig:3d-exp}
\end{figure}

%% file: framework.tex
\section{Divide-and-Learn for Performance Prediction}
\label{sec:framework}


Drawing on our observations of the configuration data, we propose \Model~--- an approach that enables better prediction of the software performance via ``divide-and-learn''. To mitigate the sample sparsity issue, the key idea of \Model~is that, since different divisions of configurations show drastically diverse characteristics, i.e., rather different performance values with distant values of key configuration options, we seek to independently learn a local model for each of those divisions that contain \textit{locally smooth} samples, thereby the learning can be more focused on the particular characteristics exhibited from the divisions and handle the feature sparsity. Yet, this requires us to formulate, on top of the original regression problem of predicting the performance value, a new classification problem without explicit labels. As such, we modify the original problem formulation (Equation~\ref{eq:prob}) as below: 
\begin{equation}
    \mathbfcal{D} = g(\mathbfcal{S}) 
\end{equation}
\begin{equation}
    \forall D_i \in \mathbfcal{D}\text{: } \mathcal{P} = f(D_i)\text{, } \mathcal{P}\in\mathbb{R}
\end{equation}
Overall, we aim to achieve three goals:

\begin{itemize}
    \item \textbf{Goal 1:} dividing the data samples into diverse yet more focused divisions $\mathbfcal{D}$ (building function $g$) and;
    \item \textbf{Goal 2:} training a dedicated local model for each division $D_i$ (building function $f$) while;
    \item \textbf{Goal 3:} assigning a newly coming configuration into the right model for prediction (using functions $g$ and $f$).
\end{itemize}

\begin{figure}[t!]
  \centering
  \includegraphics[width=\columnwidth]{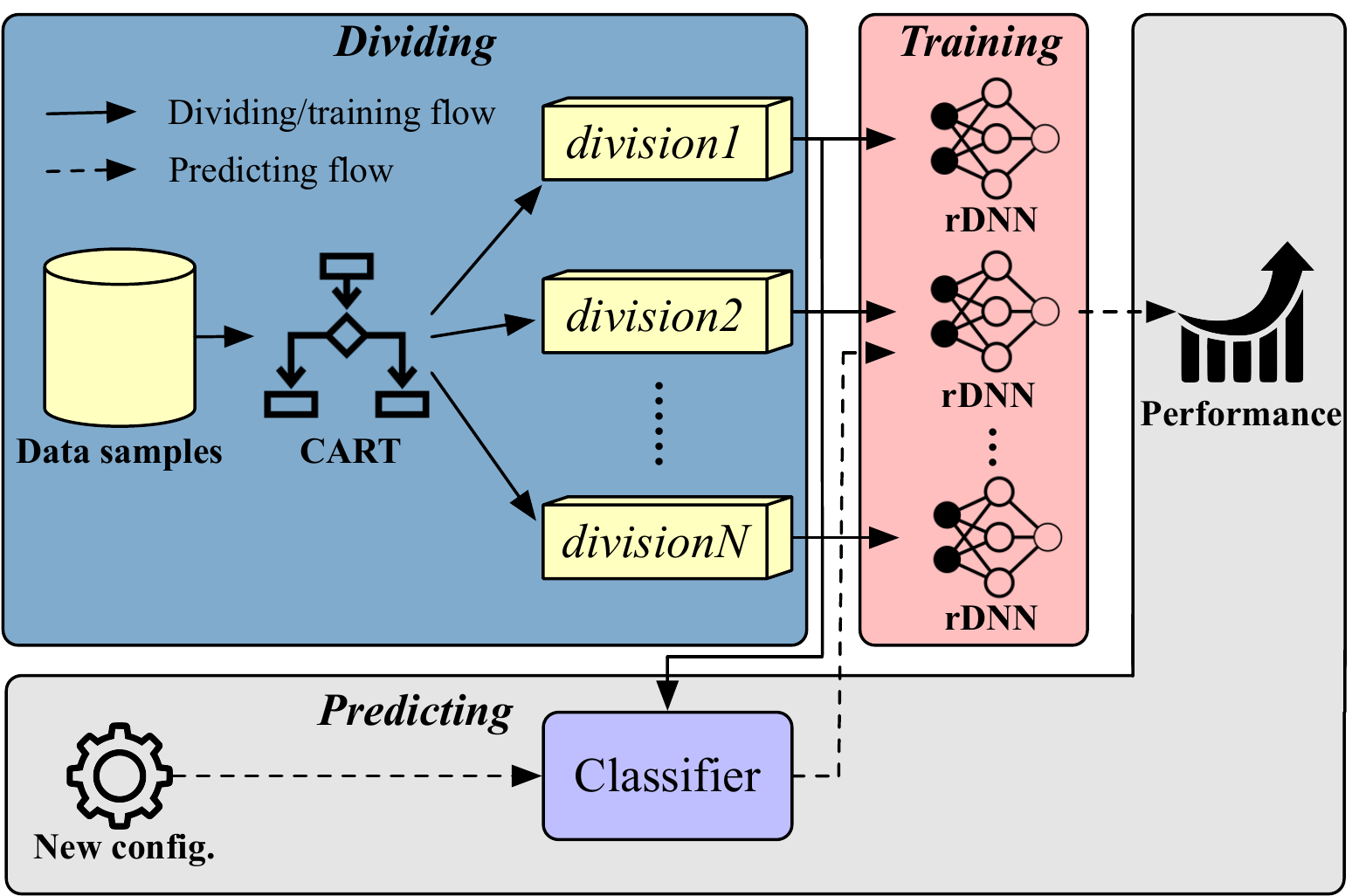}
  \caption{The architecture of \Model.}
  \label{fig:structure}
\end{figure}

Figure~\ref{fig:structure} illustrates the overall architecture of \Model, in which there are three core phases, namely \textit{Dividing}, \textit{Training}, and \textit{Predicting}. A pseudo code can also be found in Algorithm~\ref{alg:dal-code}.

\subsection{Dividing}
\label{subsec:phase1_clustering}

The very first phase in \Model~is to appropriately divide the data into more focused divisions while doing so by considering both the configuration options and performance values. To that end, the key question we seek to address is: how to effectively cluster the performance data with similar sample characteristics (\textbf{Goal 1})?

Indeed, for dividing the data samples, it makes sense to consider various unsupervised clustering algorithms, such as $k$-mean~\cite{macqueen1967some}, BIRCH~\cite{DBLP:conf/sigmod/ZhangRL96} or DBSCAN~\cite{DBLP:conf/kdd/EsterKSX96}. However, we found that they are ill-suited for our problem, because:

\begin{itemize}
    \item the distance metrics are highly system-dependent. For example, depending on the number of configuration options and whether they are binary/numeric options;
    \item it is difficult to combine the configuration options and performance value with appropriate discrimination; 
    \item and clustering algorithms are often non-interpretable.
\end{itemize}

\begin{figure}[t!]
  \centering
  \includegraphics[width=\columnwidth]{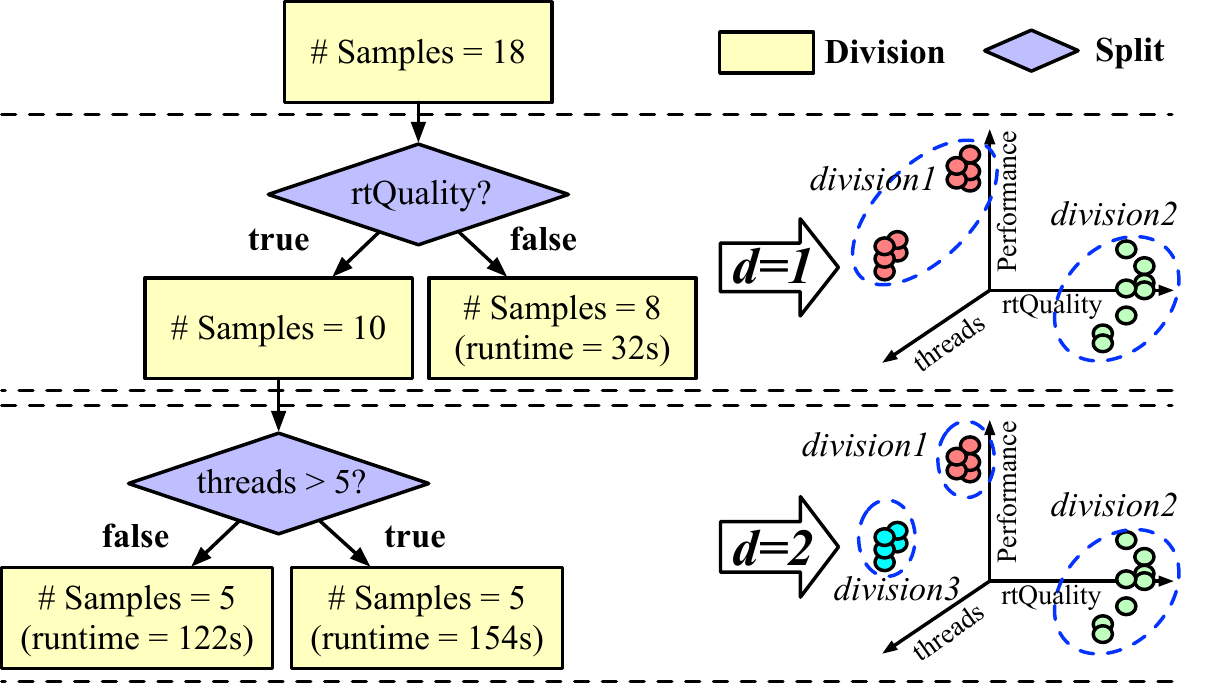}
  \caption{Projection of CART for \textsc{VP8} showing the possible divisions with different colors under alternative depth $d$.}
  \label{fig:DT_example}
\end{figure}

As a result, in \Model, we extend Classification and Regression Tree (CART) as the clustering algorithm (lines 3-11 in Algorithm~\ref{alg:dal-code}) since (1) it is simple with interpretable/analyzable structure; (2) it ranks the important options as part of training (good for dealing with the feature sparsity issue), and (3) it does not suffer the issues above~\cite{DBLP:conf/kbse/SarkarGSAC15,DBLP:journals/ese/GuoYSASVCWY18,DBLP:journals/ase/NairMSA18,DBLP:conf/icse/Chen19b,DBLP:journals/tse/ChenB17,DBLP:journals/corr/abs-1801-02175,DBLP:conf/kbse/GuoCASW13}. As illustrated in Figure~\ref{fig:DT_example}, CART is originally a supervised and binary tree-structured model, which recursively splits some, if not all, configuration options and the corresponding data samples based on tuned thresholds. A split would result in two divisions, each of which can be further split. In this work, we at first train the CART on the available samples of configurations and performance values, during which we use the most common mean performance of all samples for each division $D_i$ as the prediction~\cite{DBLP:journals/ese/GuoYSASVCWY18,DBLP:conf/kbse/GuoCASW13}:
\begin{equation}
    \overline{y}_{D_i}={{1}\over{|D_i|}} {\sum_{y_j \in D_i}y_j}
    \label{eq:seg}
\end{equation}
in which $y_j$ is a performance value. For example, Figure~\ref{fig:DT_example} shows a projected example, in which the configuration that satisfies ``\texttt{rtQua-} \texttt{lity=true}'' and ``\texttt{threads=3}'' would lead to an inferred runtime of 112 seconds, which is calculated over all the 5 samples involved using Equation~\ref{eq:seg}.

\input{Tables/alg1}

By choosing/ranking options that serve as the splits and tuning their thresholds, in \Model, we seek to minimize the following overall loss function during the CART training:
\begin{equation}
   \mathcal{L}= {1\over{|D_l|}}{\sum_{y_j \in D_l}{(y_j - \overline{y}_{D_l})}^2} + {1\over{|D_r|}}{\sum_{y_j \in D_r}{(y_j - \overline{y}_{D_r})}^2}
    \label{eq:loss}
\end{equation}
where $D_l$ and $D_r$ denote the left and right division from a split, respectively. This ensures that the divisions would contain data samples with similar performance values (\textbf{Property 1}) while they are formed with respect to the similar values of the key configuration options as determined by the splits/thresholds at the finest granularity (\textbf{Property 2}), i.e., the more important options would appear on the higher level of the tree with excessive splitting.

However, here we do not use CART to generalize prediction directly on new data once it is trained as it has been shown that the splits and simple average of performance values in the division alone can still fail to handle the complex interactions between the options, leading to insufficient accuracy~\cite{DBLP:conf/icse/HaZ19}. Further, with our loss function in Equation~\ref{eq:loss}, CART is prone to be overfitting\footnote{Overfitting means a learned model fits well with the training data but works poorly on new data.} especially for software quality data~\cite{DBLP:journals/ese/KhoshgoftaarA01}. This exacerbates the issue of sample sparsity~\cite{DBLP:journals/corr/abs-2202-03354} under a small amount of data samples which is not uncommon for configurable software systems~\cite{DBLP:conf/icse/HaZ19,DBLP:conf/esem/ShuS0X20}. 

Instead, what we are interested in are the (branch and/or leaf) divisions made therein (with respect to the training data), which enable us to use further dedicated and more focused local models for better generalizing to the new data (lines 6-11 in Algorithm~\ref{alg:dal-code}). As such, the final prediction is no longer a simple average while we do not care about the CART overfitting itself as long as it fits the training data well. This is similar to the case of unsupervised clustering for which the clustering is guided by implicit labels (via the loss function at Equation~\ref{eq:loss}). Specifically, in \Model~we extract the data samples according to the divisions made by the $d$th depth of the CART, including all the leaf divisions with depth smaller than $d$. An example can be seen from Figure~\ref{fig:DT_example}, where $d$ is a controllable parameter to be given. In this way, \Model~divides the data into a range of $[d+1,2^{d}]$ divisions ($d \geq 1$), each of which will be captured by a local model learned thereafter. Note that when the number of data samples in the division is less than the minimum amount required by a model, we merge the two divisions of the same parent node.



As a concrete example, from Figure~\ref{fig:DT_example}, we see that there are two depths: when $d=1$ there would be two divisions (one branch and one leaf) with 10 and 8 samples respectively; similarly, when $d=2$ there would be three leaf divisions: two of each have 5 samples and one is the division with 8 samples from $d=1$ as it is a leaf. In this case, CART has detected that the \texttt{rtQuality} is a more important (binary) option to impact the performance, and hence it should be considered at a higher level in the tree. Note that for numeric options, e.g., \texttt{threads}, the threshold of splitting (\texttt{threads} $>5$) is also tuned as part of the training process of CART.


\subsection{Training}

Given the divisions produced by the \textit{Dividing} phase, we train a local model for the samples from each division identified as part of \textbf{Goal 2} (lines 12-14 in Algorithm~\ref{alg:dal-code}). Theoretically, we can pair them with any model. However, as we will show in Section~\ref{subsec:rq2}, the state-of-the-art regularized Deep Neural Network (rDNN)~\cite{DBLP:conf/icse/HaZ19} (namely \texttt{DeepPerf}), published at ICSE'19, is the most effective one under \Model~as it handles feature sparsity well for configurable software. Indeed, Ha and Zhang~\cite{DBLP:conf/icse/HaZ19} showed that rDNN is more effective than the others even with small data samples when predicting software performance (in our study, we also evaluate the same systems with small training sample sizes as used in their work). Therefore, in \Model~we choose rDNN as the underlying local model by default.

In this work, we adopt exactly the same structure and training procedure as those used by Ha and Zhang~\cite{DBLP:conf/icse/HaZ19}, hence we kindly refer interested readers to their work for the training details~\cite{DBLP:conf/icse/HaZ19}. Since the local models of the divisions are independent, we utilize parallel training as part of \Model.

\subsection{Predicting}

When a new configuration arrives for prediction, \Model~chooses a model of division trained previously to infer its performance. Therefore, the question is: how to assign the new configuration to the right model (\textbf{Goal 3})? A naive solution is to directly feed the configuration into the CART from the \textit{dividing} phase and check which divisions it associates with. Yet, since the performance of the new configuration is unforeseen from the CART's training data, this solution requires CART to generalize accurately, which, as mentioned, can easily lead to poor results because CART is overfitting-prone when directly working on new data~\cite{DBLP:journals/ese/KhoshgoftaarA01}.


Instead, by using the divided samples from the \textit{Dividing} phase (which serves as pseudo labeled data), we train a Random Forest---a widely used classifier and is resilient to overfitting ~\cite{DBLP:conf/splc/ValovGC15,DBLP:conf/oopsla/QueirozBC16,DBLP:conf/icse/0003XC021}---to generalize the decision boundary and predict which division that the new configuration should be better assigned to (lines 15-21 in Algorithm~\ref{alg:dal-code}). Again, in this way, we are less concerned about the overfitting issue of CART as long as it matches the patterns of training data well. This now becomes a typical classification problem but there are only pseudo labels to be used in the training. Using the example from Figure~\ref{fig:DT_example} again, if $d=1$ then the configurations in the 10 sample set would have a label \textit{``division1''}; similarly, those in the 8 sample set would result in a label \textit{``division2''}.

However, one issue we experienced is that, even with $d=1$, the sample size of the two divisions can be rather imbalanced, which severely harms the quality of the classifier trained. For example, when training \textsc{BDB-C} with 18 samples, the first split in CART can lead to two divisions with 14 and 4 samples, respectively.
Therefore, before training the classifier we use Synthetic Minority Oversampling Technique (SMOTE)~\cite{DBLP:journals/bmcbi/BlagusL13a} to pre-process the pseudo label data, hence the division(s) with much less data (minority) can be more repeatedly sampled.


Finally, the classifier predicts a division whose local model would infer the performance of the new configuration.

\subsection{Trade-off with the Number of Divisions}
\label{sec:division}

Since more divisions mean that the sample space is separated into more loosely related regions for dealing with the sample sparsity, one may expect that the accuracy will be improved, or at least, stay similar, thereby we should use the maximum possible $d$ from CART in the \textit{dividing phase}. This, however, only exists in the ``utopia case'' where there is an infinite set of configuration data samples.

In essence, with the design of \Model, the depth $d$ will manage two conflicting goals that influence its accuracy:

\begin{enumerate}
    \item greater ability to handle sample sparsity by separating the distant samples into divisions, each of which is learned by an isolated local model;
    \item and a larger amount of data samples in each division for the local model to be able to generalize.
\end{enumerate}

Clearly, a greater $d$ may benefit goal (1) but it will inevitably damage goal (2) since it is possible for CART to generate divisions with imbalanced sample sizes. As a result, we see $d$ as a value that controls the trade-off between the two goals, and neither a too small nor too large $d$ would be ideal, as the former would lose the ability to deal with sample sparsity while the latter would leave too little data for a local model to learn, hence produce negative noises to the overall prediction. Similar to the fact that we cannot theoretically justify how much data is sufficient for a model to learn the concept~\cite{DBLP:conf/dyspan/OyedareP19}, it is also difficult to prove how many divisions are sufficient for handling the sample sparsity in performance modeling. However, in Section~\ref{subsec:rq3}, we will empirically demonstrate that there is a (upward) quadratic correlation between $d$ value and the error incurred by \Model~due to the conflict between the above two goals.


%% file: Tables/alg1.tex
\begin{algorithm}[t!]
	\DontPrintSemicolon
	\footnotesize
	
	\caption{Pseudo code of \Model}
	\label{alg:dal-code}
	\KwIn{The expected depth $d$ extracted from CART; a new configuration $\mathbf{\overline{c}}$ to be predicted}
     \KwOut{The predicted performance of $\mathbf{\overline{c}}$}
	
	\If{$\mathbfcal{M} = \emptyset$}
	{
	  \tcc{\textcolor{blue}{dividing phase.}}
      $\mathbfcal{S}\leftarrow$ randomly sample a set of configurations and their performance\\
      $\mathcal{T}\leftarrow$ \textsc{trainCART($\mathbfcal{S}$)}\\
      $d'=1$\\
      \While{$d'\leq d$}{
           \If{$d'<d$}
           {
             $\mathbfcal{D}\leftarrow$ extract all the leaf divisions of samples from $\mathcal{T}$ at the $d'$th depth\\
           }
           \Else{
              $\mathbfcal{D}\leftarrow$ extract all divisions of samples from $\mathcal{T}$ at the $d'$th depth\\
           }
           $d'=d'+1$\\
      }
      
      \tcc{\textcolor{blue}{training phase.}}
      \For{$D_i \in \mathbfcal{D}$}  
      {
       $\mathbfcal{M}\leftarrow$ \textsc{trainRegularizedDNN($D_i$)}\\
    
      }
    
    }
       
    \tcc{\textcolor{blue}{predicting phase.}}
    \If{$\mathcal{F}$ has not been trained}
	{
	$\mathbfcal{U}\leftarrow$ Removing performance data and labeling the configurations based on their divisions in $\mathbfcal{D}$\\
	$\mathbfcal{U'}\leftarrow$ \textsc{SMOTE($\mathbfcal{U}$)}\\
	$\mathcal{F}\leftarrow$ \textsc{trainRandomForest($\mathbfcal{U'}$)}
	}
	
    $D_i=$  \textsc{predict($\mathcal{F}$,$\mathbf{\overline{c}}$)}\\
    $\mathcal{M}=$ get the model from $\mathbfcal{M}$ that corresponds to the predicted division $D_i$\\

    \Return \textsc{predict($\mathcal{M}$,$\mathbf{\overline{c}}$)}\\
	
\end{algorithm}

%% file: study.tex
\section{Experiment Setup}
\label{sec:setup}

Here, we delineate the settings of our evaluation. In this work, \Model~is implemented based on Tensorflow and \texttt{scikit-learn}. All experiments were carried out on a machine with Intel Core i7 2GHz CPU and 16GB RAM. 

\begin{table}[t!]

\centering
\footnotesize
\caption{Details of the subject systems. ($|\mathbfcal{B}|$/$|\mathbfcal{N}|$) denotes the number of binary/numerical options, and $|\mathbfcal{C}|$ denotes the number of valid configurations (full sample size).}
\begin{adjustbox}{width=\columnwidth,center}
\input{Tables/subject_systems}
\end{adjustbox}
\label{tb:subject-system}
\end{table}

\subsection{Research Questions}

In this work, we comprehensively assess \Model~by answering the following research questions (RQ):

\begin{itemize}
    \item \textbf{RQ1:} How accurate is \Model~compared with the state-of-the-art approaches for software performance prediction?

    \item \textbf{RQ2:} Can \Model~benefit different models when they are used locally therein for predicting software performance?

    \item \textbf{RQ3:} What is the sensitivity of \Model's accuracy to $d$?
    
    \item \textbf{RQ4:} What is the model building time for \Model?

\end{itemize}

We ask \textbf{RQ1} to assess the effectiveness of \Model~under different sample sizes against the state-of-the-art. Since the default rDNN in \Model~is replaceable, we study \textbf{RQ2} to examine how the concept of ``divide-and-learn'' can benefit any given local model and whether using rDNN as the underlying local model is the best option. In \textbf{RQ3}, we examine how the depth of division ($d$) can impact the performance of \Model. Finally, we examine the overall overhead of \Model~in \textbf{RQ4}.

\subsection{Subject Systems}
\label{subsec:subject_system}
We use the same datasets of all valid configurations from real-world systems as widely used in the literature~\cite{DBLP:conf/esem/ShuS0X20,DBLP:conf/icse/HaZ19,DBLP:conf/sigsoft/SiegmundGAK15, DBLP:journals/ese/GuoYSASVCWY18, DBLP:journals/corr/abs-2106-02716}. To reduce noise, we remove those that contain missing measurements or invalid configurations. As shown in Table~\ref{tb:subject-system}, we consider eight configurable software systems with diverse domains, scales, and performance concerns. Some of those contain only binary configuration options (e.g., \textsc{x264}) while the others involve mixed options (binary and numeric), e.g., \textsc{HSMGP}, which can be more difficult to model~\cite{DBLP:conf/icse/HaZ19}.


The configuration data of all the systems are collected by prior studies using the standard benchmarks with repeated measurement~\cite{DBLP:conf/esem/ShuS0X20,DBLP:conf/icse/HaZ19,DBLP:conf/sigsoft/SiegmundGAK15, DBLP:journals/ese/GuoYSASVCWY18, DBLP:journals/corr/abs-2106-02716}. For example, the configurations of \textsc{Apache}---a popular Web server---are benchmarked using the tools \texttt{Autobench} and \texttt{Httperf}, where workloads are generated and increased until reaching the point before the server crashes, and then the maximum load is marked as the performance value~\cite{DBLP:journals/ese/GuoYSASVCWY18}. The process repeats a few times for each configuration to ensure reliability.


To ensure generalizability of the results, for each system, we follow the protocol used by existing work~\cite{DBLP:journals/sqj/SiegmundRKKAS12,DBLP:conf/icse/HaZ19,DBLP:conf/esem/ShuS0X20} to obtain five sets of training sample size in the evaluation: 

\begin{itemize}
    \item \textbf{Binary systems:} We randomly sample $n$, $2n$, $3n$, $4n$, and $5n$ configurations and their measurements, where $n$ is the number of configuration options~\cite{DBLP:conf/icse/HaZ19,DBLP:conf/esem/ShuS0X20}.
    \item \textbf{Mixed systems:} We leverage the sizes suggested by \texttt{SPLCon}-\texttt{queror}~\cite{DBLP:journals/sqj/SiegmundRKKAS12} (a state-of-the-art approach) depending on the amount of budget. 
    
\end{itemize}

The results have been illustrated in Table~\ref{tb:sizes}. All the remaining samples in the dataset are used for testing.

\input{Tables/sample_sizes}

\subsection{Metric and Statistical Validation}

\subsubsection{Accuracy}

For all the experiments, mean relative error (MRE) is used as the evaluation metric for prediction accuracy, since it provides an intuitive indication of the error and has been widely used in the domain of software performance prediction~\cite{DBLP:conf/icse/HaZ19, DBLP:conf/esem/ShuS0X20, DBLP:journals/ese/GuoYSASVCWY18}. Formally, the MRE is computed as:
\begin{equation}
     MRE = {{1} \over {k}} \times {\sum^k_{t=1} {{|A_t - P_t|} \over {A_t}}} \times 100\%
\end{equation}
\noindent whereby $A_t$ and $P_t$ denote the $t$th actual and predicted performance, respectively. To mitigate bias, all experiments are repeated for 30 runs via bootstrapping without replacement. Note that excluding replacement is a common strategy for the performance learning of configuration as it is rare for a model to learn from the same configuration sample more than once~\cite{DBLP:conf/msr/GongChen22}.

\subsubsection{Statistical Test}




    
    




Since our evaluation commonly involves comparing more than two approaches, we apply Scott-Knott test~\cite{DBLP:journals/tse/MittasA13} to evaluate their statistical significance on the difference of MRE over 30 runs, as recommended by Mittas and Angelis~\cite{DBLP:journals/tse/MittasA13}. In a nutshell, Scott-Knott sorts the list of treatments (the approaches that model the system) by their median values of the MRE. Next, it splits the list into two sub-lists with the largest expected difference~\cite{xia2018hyperparameter}. For example, suppose that we compare $A$, $B$, and $C$, a possible split could be $\{A, B\}$, $\{C\}$, with the rank ($r$) of 1 and 2, respectively. This means that, in the statistical sense, $A$ and $B$ perform similarly, but they are significantly better than $C$. Formally, Scott-Knott test aims to find the best split by maximizing the difference $\Delta$ in the expected mean before and after each split:
\begin{equation}
    \Delta = \frac{|l_1|}{|l|}(\overline{l_1} - \overline{l})^2 + \frac{|l_2|}{|l|}(\overline{l_2} - \overline{l})^2
\end{equation}
whereby $|l_1|$ and $|l_2|$ are the sizes of two sub-lists ($l_1$ and $l_2$) from list $l$ with a size $|l|$. $\overline{l_1}$, $\overline{l_2}$, and $\overline{l}$ denote their mean MRE.

During the splitting, we apply a statistical hypothesis test $H$ to check if $l_1$ and $l_2$ are significantly different. This is done by using bootstrapping and $\hat{A}_{12}$~\cite{Vargha2000ACA}. If that is the case, Scott-Knott recurses on the splits. In other words, we divide the approaches into different sub-lists if both bootstrap sampling and effect size test suggest that a split is statistically significant (with a confidence level of 99\%) and with a good effect $\hat{A}_{12} \geq 0.6$. The sub-lists are then ranked based on their mean MRE. 



%% file: Tables/subject_systems.tex
\footnotesize
\setlength{\tabcolsep}{1mm}
\begin{tabular}{cccccc}
\toprule
\textbf{System} & \textbf{$|\mathbfcal{B}|$/$|\mathbfcal{N}|$} & \textbf{Performance} & \textbf{Description} & \textbf{$|\mathbfcal{C}|$} & \textbf{Used by}\\ 
\midrule
\textsc{Apache} & 9/0  & Maximum load & Web server              & 192 & \cite{DBLP:conf/icse/HaZ19,DBLP:journals/ese/GuoYSASVCWY18,DBLP:conf/esem/ShuS0X20} \\
\textsc{BDB-C}  & 16/0 & Latency (ms) & Database (C)    & 2560 & \cite{DBLP:conf/icse/HaZ19,DBLP:journals/ese/GuoYSASVCWY18,DBLP:conf/esem/ShuS0X20} \\
\textsc{BDB-J}  & 26/0 & Latency (ms) & Database (Java) & 180 & \cite{DBLP:conf/icse/HaZ19,DBLP:journals/ese/GuoYSASVCWY18,DBLP:conf/esem/ShuS0X20} \\
\textsc{x264}   & 16/0 & Runtime (ms)  & Video encoder           & 1152 & \cite{DBLP:conf/icse/HaZ19,DBLP:journals/ese/GuoYSASVCWY18,DBLP:conf/esem/ShuS0X20} \\
\textsc{HSMGP}   & 11/3 & Runtime (ms) & Compiler                & 3456 & \cite{DBLP:conf/icse/HaZ19,DBLP:conf/sigsoft/SiegmundGAK15} \\
\textsc{HIPA$^{cc}$}   & 31/2 & Runtime (ms) & Compiler                & 13485 & \cite{DBLP:conf/icse/HaZ19,DBLP:conf/sigsoft/SiegmundGAK15} \\
\textsc{VP8}   & 9/4 & Runtime (ms) & Video encoder                & 2736 & \cite{DBLP:journals/corr/abs-2106-02716} \\
\textsc{Lrzip}   & 9/3 & Runtime (ms) & Compression tool                & 5184 & \cite{DBLP:journals/corr/abs-2106-02716} \\
\bottomrule
\end{tabular}

%% file: Tables/sample_sizes.tex
\begin{table}[t!]
\caption{The training sample sizes used. $n$ denotes the number of configuration options in a binary system.}
\centering
\footnotesize
\begin{adjustbox}{width=0.75\linewidth,center}
\begin{tabular}{cccccc}
\toprule
\textbf{System} & \textbf{Size 1} & \textbf{Size 2} & \textbf{Size 3} & \textbf{Size 4} & \textbf{Size 5} \\ 
\midrule
\textsc{Apache} & $n$  & $2n$ & $3n$  & $4n$ & $5n$\\
\textsc{BDB-C}  & $n$  & $2n$ & $3n$  & $4n$ & $5n$\\
\textsc{BDB-J}  & $n$  & $2n$ & $3n$  & $4n$ & $5n$\\
\textsc{x264}   & $n$  & $2n$ & $3n$  & $4n$ & $5n$\\
\textsc{HSMGP} & 77 & 173 & 384 & 480 & 864\\
\textsc{HIPA$^{cc}$}  & 261 & 528 & 736 & 1281 & 2631\\
\textsc{VP8} & 121 & 273 & 356 & 467 & 830 \\
\textsc{Lrzip} & 127 & 295 & 386 & 485 & 907\\
\bottomrule
\end{tabular}
\end{adjustbox}
\label{tb:sizes}
\end{table}

%% file: evaluation.tex
\begin{table*}[t!]
\footnotesize
\setlength{\tabcolsep}{1mm}
\caption{The median and interquartile range of MRE, denoted as Med (IQR), for \Model~and the state-of-the-art approaches for all the subject systems and training sizes over 30 runs. For each case, \setlength{\fboxsep}{1.5pt}\colorbox{green!20}{green cells} mean \Model~has the best median MRE; or \setlength{\fboxsep}{1.5pt}\colorbox{red!20}{red cells} otherwise. The one(s) with the best rank ($r$) from the Scott-Knott test is highlighted in bold.}
\begin{adjustbox}{width=\textwidth,center}
\input{Tables/vs_STOA_scottknott}
\end{adjustbox}
\label{tb:vsSOTA}
\end{table*}

\section{Evaluation}
\label{sec:evaluation}


\subsection{Comparing with the State-of-the-art}
\label{subsec:rq1}

\subsubsection{Method}

To understand how \Model~performs compared with the state-of-the-art, we assess its accuracy against both the standard approaches that rely on statistical learning, i.e., \texttt{SPLConqueror}~\cite{DBLP:journals/sqj/SiegmundRKKAS12} (linear regression and sampling methods) and \texttt{DECART}~\cite{DBLP:journals/ese/GuoYSASVCWY18} (an improved CART), together with recent deep learning-based ones, i.e., \texttt{DeepPerf}~\cite{DBLP:conf/icse/HaZ19} (a single global rDNN) and \texttt{Perf-AL}~\cite{DBLP:conf/esem/ShuS0X20} (an adversarial learning method). All approaches can be used for any type of system except for \texttt{DECART}, which works on binary systems only. Following the setting used by Ha and Zhang~\cite{DBLP:conf/icse/HaZ19},  \texttt{SPLConqueror}\footnote{Since \texttt{SPLConqueror} supports multiple sampling methods, we use the one (or combination for the mixed system) that leads to the best MRE.} and \texttt{DECART} use their own sampling method while \Model, \texttt{DeepPerf} and \texttt{Perf-AL} rely on random sampling. Since there are 8 systems and 5 sample sizes each, we obtain 40 cases to compare in total.


We use the implementations published by their authors with the same parameter settings. For \Model, we set $d=1$ or $d=2$ depending on the systems, which tends to be the most appropriate value based on the result under a small portion of training data (see Section~\ref{subsec:rq3}). We use the systems, training sizes, and statistical tests as described in Section~\ref{sec:setup}. All experiments are repeated for 30 runs. 

\subsubsection{Results}

The results have been illustrated in Table~\ref{tb:vsSOTA}, from which we see that \Model~remarkably achieves the best accuracy on 31 out of 40 cases. In particular, \Model~considerably improves the accuracy, e.g., by up to $1.94\times$ better than the second-best one on \textit{Size 1} of \textsc{VP8}. The above still holds when looking into the results of the statistical test: \Model~is the only approach that is ranked first for 26 out of the 31 cases. For the 9 cases where \Model~does not achieve the best median MRE, it is equally ranked as the first for two of them. These conclude that \Model~is, in 33 cases, similar to (7 cases) or significantly better (26 cases) than the best state-of-the-art for each specific case (which could be a different approach). 

For cases with different training sample sizes, we see that \Model~performs generally more inferior than the others when the size is too limited, i.e., \textit{Size 1} and \textit{Size 2} for the binary systems. This is expected as when there are too few samples, each local model would have a limited chance to observe the right pattern after the splitting, hence blurring its effectiveness in handling sample sparsity. However, in the other cases (especially for mixed systems that have more data even for \textit{Size 1}), \Model~needs far fewer samples to achieve the same accuracy as the best state-of-the-art. For example, on \textsc{Lrzip}, \Model~only needs 386 samples (\textit{Size 3}) to achieve an error less than 15\% while \texttt{DeepPerf} requires 907 samples (\textit{Size 5}) to do so.

Another observation is that the improvements of \Model~is much more obvious in mixed systems than those for binary systems. This is because: (1) the binary systems have fewer training samples as they have a smaller configuration space. Therefore, the data learned by each local model is more restricted. (2) The issue of sample sparsity is more severe on mixed systems, as their configuration landscape is more complex and comes with finer granularity.


As a result, we anticipate that the benefit of \Model~can be amplified with more complex systems and/or more training data.



\begin{table*}[h!]
\centering
\footnotesize
\caption{The Scott-Knott ranks ($r$) on the MRE of \Model~under different local models and their global counterparts. The \colorbox{green!20}{green cells} denote the best rank. Raw MRE results can be accessed at:  \texttt{\textcolor{blue}{\protect\url{https://github.com/ideas-labo/DaL/blob/main/Table4_full.pdf}}}.}
\begin{adjustbox}{width=\textwidth,center}
\input{Tables/compare_models}
\end{adjustbox}
\label{tb:scott-knott}
\end{table*}


To summarize, we can answer \textbf{RQ1} as:

\begin{quotebox}
   \noindent
   \textit{\textbf{RQ1:} \Model~performs similar or significantly better than the best state-of-the-art approach in 33 out of 40 cases, with up to $1.94\times$ improvements. It also needs fewer samples to achieve the same accuracy and the benefits can be amplified with complex systems/more training samples.} 
\end{quotebox}

\subsection{\Model~under Different Local Models}
\label{subsec:rq2}

\subsubsection{Method}

Since the idea of ``divide-and-learn'' can be applicable to a wide range of underlying local models of the divisions identified, we seek to understand how well \Model~perform with different local models against their global model counterparts (i.e., using them directly to learn the entire training dataset). To that end, we run experiments on a set of global models available in \texttt{scikit-learn} and widely used in software engineering tasks to make predictions directly~\cite{DBLP:conf/icse/LiX0WT20,DBLP:conf/msr/GongChen22}, such as CART, Random Forest (RF), Linear Regression (LR),  Support Vector Regression (SVR), Kernel Ridge Regression (KRR), and $k$-Nearest Neighbours ($k$NN). We used the same settings as those for \textbf{RQ1} and all models' hyperparameters are tuned in training. For the simplicity of exposition, we report on the ranks $r$ produced by the Scott-Knott test. 


\subsubsection{Result}

From Table~\ref{tb:scott-knott}, we can obtain the following key observations: firstly, when examining each pair of the counterparts, i.e., \texttt{DaL$_X$} and \textit{X}, \Model~can indeed improve the accuracy of the local model via the concept of ``divide-and-learn''. In particular, for simple but commonly ineffective models like LR~\cite{DBLP:conf/icse/HaZ19}, \Model~can improve them to a considerable extent. Yet, we see that \Model~does not often lead to a significantly different result when working with CART against using CART directly. This is as expected, since using different CART models for the divisions identified by a CART makes little difference to applying a single CART that predicts directly. Interestingly, we also see that our model performs better than the traditional ensemble learning: \texttt{DaL$_{CART}$}---a CART-based ``divide-and-learn'' model performs generally better than RF, which uses CART as the local model and combines them via Bagging.

Secondly, the default of \Model, which uses the rDNN as the local model, still performs significantly better than the others. This aligns with the findings from existing work~\cite{DBLP:conf/icse/HaZ19} that the rDNN handles the feature sparsity better. Indeed, deep learning models are known to be data-hungry, but our results surprisingly show that they can also work well for a limited amount of configuration samples. The key behind such is the use of regularization, which stresses additional penalties on the more important weights/options. This has helped to relieve the need for a large amount of data during training while better fitting with the sparse features in configuration data. A similar conclusion has also been drawn from previous studies~\cite{DBLP:conf/icse/HaZ19}.

Therefore, for \textbf{RQ2}, we say:

    
    
    
    

\begin{quotebox}
   \noindent
   \textit{\textbf{RQ2:}} Thanks to the concept of ``divide-and-learn'', \Model~is able to significantly improve a range of global models when using them as the underlying local model.
\end{quotebox}

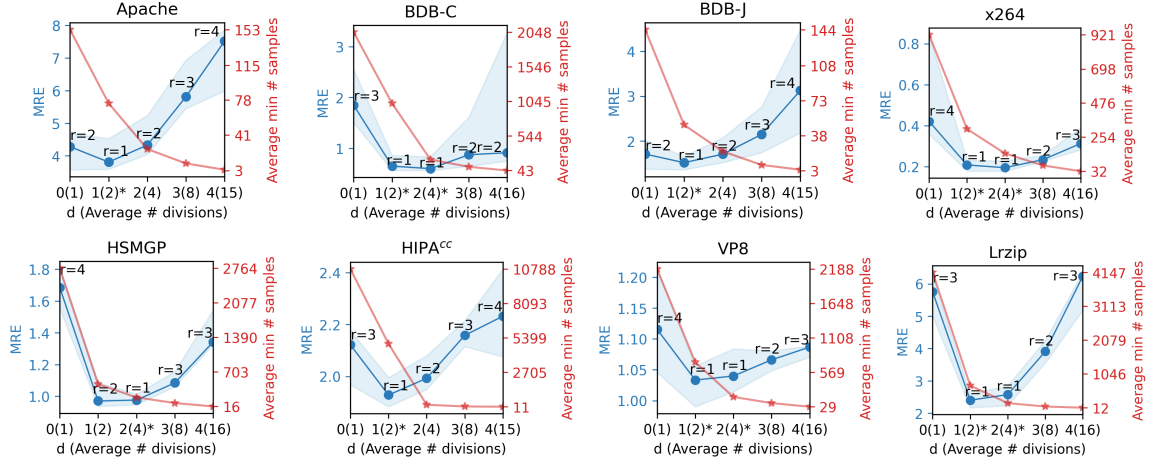
\begin{figure*}[!t]
    \centering
    \footnotesize
    \input{Figures/Sensitivity_to_depth}
    \caption{The median MRE (\markone{0}{20}{10}{20}), its IQR (area), and the average smallest training size of the divisions (\marktwo{0}{20}{7.5}{20}) achieved by \Model~under different depths ($d$ values)/number of divisions over all systems and 30 runs. $r=1$ means rank 1 in the Scott-Knott test on MRE. The best-ranked $d$ is marked as $*$.}
    \label{fig:depths}
\end{figure*}


\subsection{Sensitivity to the Depth $d$}
\label{subsec:rq3}

\subsubsection{Method}
To understand \textbf{RQ3}, we examine different $d$ values. Since the number of divisions (and hence the possible depth) is sample size-dependent, for each system, we use 80\% of the full dataset for training and the remaining for testing. This has allowed us to achieve up to $d=4$ with 16 divisions as the maximum possible bound. For different $d$ values, we report on the median MRE together with the results of Scott-Knott test for 30 runs. We also report the smallest sample size from the divisions, averaging over 30 runs.

\subsubsection{Results}

From Figure~\ref{fig:depths}, we see that the correlation between the error of \Model~and $d$ value is close to quadratic: \Model~reaches its best MRE with $d=1$ or $d=2$. At the same time, the size of training data for a local model decreases as the number of divisions increases. Since $d$ controls the trade-off between the ability to handle sample sparsity and ensuring sufficient data samples to train all local models, $d=1$ or $d=2$ tends to be the ``sweet points'' that reach a balance for the systems studied. After the point of $d=1$ or $d=2$, the MRE will worsen, as the local models' training size often drops dramatically. This is a clear sign that, from that point, the side-effect of having too less samples to train a local model has started to surpass the benefit that could have been brought by dealing with sample sparsity using more local models.

When $d=0$, which means only one division and hence \Model~is reduced to \texttt{DeepPerf} that ignores sample sparsity, the resulted MRE is the worst on 4 out of 8 systems; the same applied to the case when $d=4$. This suggests that neither too small $d$ (e.g., $d=0$ with only one division) nor too larger $d$ (e.g., $d=4$ with up to 16 divisions, i.e., too many divisions) are ideal, which matches our theoretical analysis in Section~\ref{sec:division}.

Therefore, we conclude that:

\begin{quotebox}
   \noindent
   \textit{\textbf{RQ3:} The error of \Model~has a (upward) quadratic correlation to $d$. In this work, $d=1$ or $d=2$ (2 to 4 divisions) reaches a good balance between handling sample sparsity and providing sufficient training data for the local models.}
   
\end{quotebox}

\subsection{Overhead of Model Building}
\label{subsec:rq4}


\subsubsection{Method}

To study \textbf{RQ4}, we examine the overall time required and the breakdown of overhead for \Model~in various phases. As some baselines, we also illustrate the model building time required by the approaches compared in \textbf{RQ1}.

\subsubsection{Result}

\input{Tables/min_max_time}

From Table~\ref{tb:max_min}, \Model~incurs an overall overhead from 6 to 56 minutes. Yet, from the breakdown, we note that the majority of the overhead comes from the \textit{training phase} that trains the local models. This is expected, as \Model~uses rDNN by default. 

Specifically, the overhead of \Model~compared with \texttt{DeepPerf} (3 to 60 minutes) is encouraging as it tends to be faster in the worst-case scenario while achieving up to $1.94\times$ better accuracy. This is because (1) each local model has less data to train and (2) the parallel training indeed speeds up the process. In contrast to \texttt{Perf-AL} (a few seconds to one minute), \Model~appears to be rather slow as the former does not use hyperparameter tuning but fixed-parameter values~\cite{DBLP:conf/esem/ShuS0X20}. Yet, as we have shown for \textbf{RQ1}, \Model~achieves up to a few magnitudes of accuracy improvement. Although \texttt{SPLConqueror} and \texttt{DECART} have an overhead of less than a minute, again their accuracy is much more inferior. Further, \texttt{SPLConqueror} requires a good selection of the sampling method(s) (which can largely incur additional overhead) while \texttt{DECART} does not work on mixed systems. Finally, we have shown in \textbf{RQ3} that \Model's MRE is quadratically sensitive to $d$ (upward), hence its value should be neither too small nor too large, e.g., $d=1$ or $d=2$ in this work.


In summary, we say that:

\begin{quotebox}
   \noindent
   \textit{\textbf{RQ4:} \Model~has competitive model building time to \texttt{DeepPerf} and higher overhead than the other state-of-the-art approaches, but this can be acceptable considering its improvement in accuracy.}
\end{quotebox}





%% file: Tables/vs_STOA_scottknott.tex
\resizebox{\textwidth}{!}{ 
\begin{tabular}{p{0.7cm}p{2cm}p{0.3cm}p{1.55cm}|p{0.3cm}p{1.55cm}|p{0.3cm}p{1.4cm}|p{0.3cm}p{1.4cm}|p{0.3cm}p{1.55cm}|p{0.3cm}p{1.4cm}|p{0.3cm}p{1.4cm}|p{0.3cm}p{1.55cm}}
\toprule
 &
  \multirow{2}{*}{\textbf{Approach}} &
  \multicolumn{2}{c}{\textsc{\textbf{Apache}}} &
  \multicolumn{2}{c}{\textsc{\textbf{BDB-C}}} &
  \multicolumn{2}{c}{\textsc{\textbf{BDB-J}}} &
  \multicolumn{2}{c}{\textsc{\textbf{x264}}} &
  \multicolumn{2}{c}{\textsc{\textbf{HSMGP}}} &
  \multicolumn{2}{c}{\textsc{\textbf{HIPA$^{cc}$}}} &
  \multicolumn{2}{c}{\textsc{\textbf{VP8}}} &
  \multicolumn{2}{c}{\textsc{\textbf{Lrzip}}} \\ \cline{3-18}

 &
   &
  \textbf{$r$} &
  \textbf{Med (IQR)} &
  \textbf{$r$} &
  \textbf{Med (IQR)} &
  \textbf{$r$} &
  \textbf{Med (IQR)} &
  \textbf{$r$} &
  \textbf{Med (IQR)} &
  \textbf{$r$} &
  \textbf{Med (IQR)} &
  \textbf{$r$} &
  \textbf{Med (IQR)} &
  \textbf{$r$} &
  \textbf{Med (IQR)} &
  \textbf{$r$} &
  \textbf{Med (IQR)} \\ \hline
 &
  \texttt{DeepPerf} &
  3 &
  20.19 (6.34) &
  \textbf{1} &
  \textbf{43.66 (42.88)} &
  3 &
  2.98 (3.34) &
  2 &
  8.04 (3.06) &
  3 &
  7.09 (3.04) &
  2 &
  9.70 (1.28) &
  2 &
  4.68 (3.27) &
  2 &
  35.40 (16.59) \\
 &
  \texttt{DECART} &
  2 &
  19.44 (6.48) &
  3 &
  51.88 (45.91) &
  \cellcolor{red!20}\textbf{1} &
  \cellcolor{red!20}\textbf{2.36 (0.75)} &
  3 &
  9.27 (2.11) &
  N/A &
  N/A &
  N/A &
  N/A &
  N/A &
  N/A &
  N/A &
  N/A \\
 &
  \texttt{Perf-AL} &
  4 &
  33.57 (11.34) &
  4 &
  82.38 (57.27) &
  5 &
  37.45 (2.45) &
  4 &
  37.00 (9.43) &
  4 &
  66.70 (14.05) &
  4 &
  31.99 (0.06) &
  4 &
  60.05 (2.03) &
  3 &
  58.45 (0.12) \\
 &
  \texttt{SPLConqueror} &
  \cellcolor{red!20}\textbf{1} &
  \cellcolor{red!20}\textbf{14.44 (0.00)} &
  4 &
  86.85 (0.00) &
  4 &
  12.29 (0.00) &
  \cellcolor{red!20}\textbf{1} &
  \cellcolor{red!20}\textbf{7.01 (0.00)} &
  2 &
  15.04 (0.00) &
  3 &
  16.68 (0.00) &
  3 &
  29.39 (0.00) &
  4 &
  123.7 (0.00) \\
\multirow{-5}{*}{Size 1} &
  \Model &
  2 &
  21.02 (6.64) &
  \cellcolor{green!20}\textbf{1} &
  \cellcolor{green!20}\textbf{41.77 (32.54)} &
  2 &
  3.14 (2.90) &
  2 &
  8.04 (1.30) &
  \cellcolor{green!20}\textbf{1} &
  \cellcolor{green!20}\textbf{4.66 (1.32)} &
  \cellcolor{green!20}\textbf{1} &
  \cellcolor{green!20}\textbf{9.07 (1.18)} &
  \cellcolor{green!20}\textbf{1} &
  \cellcolor{green!20}\textbf{1.59 (0.36)} &
  \cellcolor{green!20}\textbf{1} &
  \cellcolor{green!20}\textbf{26.40 (6.94)} \\ \hline
 &
  \texttt{DeepPerf} &
  \textbf{1} &
  \textbf{9.79 (4.56)} &
  2 &
 18.17 (8.11) &
  2 &
  1.91 (0.68) &
  \cellcolor{red!20}\textbf{1} &
  \cellcolor{red!20}\textbf{3.17 (1.00)} &
  2 &
  3.69 (0.61) &
  2 &
  6.89 (1.44) &
  2 &
  2.39 (1.21) &
  2 &
  21.46 (4.53) \\
 &
  \texttt{DECART} &
  \cellcolor{red!20}2 &
  \cellcolor{red!20}9.77 (5.36) &
  \cellcolor{red!20}\textbf{1} &
  \cellcolor{red!20}\textbf{13.37 (7.25)} &
  \cellcolor{red!20}\textbf{1} &
  \cellcolor{red!20}\textbf{1.84 (0.17)} &
  2 &
  6.29 (1.72) &
  N/A &
  N/A &
  N/A &
  N/A &
  N/A &
  N/A &
  N/A &
  N/A \\
 &
  \texttt{Perf-AL} &
  4 &
  32.97 (6.24) &
  4 &
  73.99 (22.51) &
  5 &
  37.90 (1.67) &
  4 &
  34.98 (7.81) &
  4 &
  66.67 (0.11) &
  4 &
  31.98 (0.06) &
  4 &
  60.04 (0.24) &
  3 &
  58.46 (0.18) \\
 &
  \texttt{SPLConqueror} &
  3 &
  13.88 (0.00) &
  2 &
  162.91 (0.00) &
  4 &
  24.85 (0.00) &
  3 &
  6.63 (0.00) &
  3 &
  22.63 (0.00) &
  3 &
  16.96 (0.00) &
  3 &
  35.34 (0.00) &
  4 &
  171.85 (0.00) \\
\multirow{-5}{*}{Size 2} &
  \Model &
  2 &
  9.82 (5.39) &
  3 &
  17.22 (15.67) &
  3 &
  1.90 (0.37) &
  \textbf{1} &
  \textbf{3.21 (1.95)} &
  \cellcolor{green!20}\textbf{1} &
  \cellcolor{green!20}\textbf{2.66 (0.68)} &
  \cellcolor{green!20}\textbf{1} &
  \cellcolor{green!20}\textbf{5.55 (0.60)} &
  \cellcolor{green!20}\textbf{1} &
  \cellcolor{green!20}\textbf{1.21 (0.09)} &
  \cellcolor{green!20}\textbf{1} &
  \cellcolor{green!20}\textbf{15.37 (6.04)} \\ \hline
 &
  \texttt{DeepPerf} &
  2 &
  7.98 (1.91) &
  2 &
  12.47 (4.08) &
  3 &
  2.01 (1.10) &
  2 &
  2.23 (0.90) &
  2 &
  2.28 (0.37) &
  2 &
  4.68 (0.90) &
  2 &
  1.93 (0.72) &
  2 &
  18.68 (3.20) \\
 &
  \texttt{DECART} &
  2 &
  8.07 (1.05) &
  \textbf{1} &
  \textbf{7.56 (5.07)} &
  \textbf{1} &
  \textbf{1.64 (0.24)} &
  3 &
  4.57 (1.03) &
  N/A &
  N/A &
  N/A &
  N/A &
  N/A &
  N/A &
  N/A &
  N/A \\
 &
  \texttt{Perf-AL} &
  4 &
  31.73 (4.66) &
  2 &
  69.74 (4.33) &
  5 &
  37.19 (1.70) &
  5 &
  36.25 (7.73) &
  4 &
  66.63 (0.18) &
  4 &
  31.99 (0.06) &
  4 &
  60.02 (0.29) &
  3 &
  58.44 (0.21) \\
 &
  \texttt{SPLConqueror} &
  3 &
  13.99 (0.00) &
  3 &
  193.89 (0.00) &
  4 &
  25.88 (0.00) &
  4 &
  6.31 (0.00) &
  3 &
  32.46 (0.00) &
  3 &
  17.18 (0.00) &
  3 &
  36.26 (0.00) &
  4 &
  177.61 (0.00) \\
\multirow{-5}{*}{Size 3} &
  \Model &
  \cellcolor{green!20}\textbf{1} &
  \cellcolor{green!20}\textbf{7.17 (2.67)} &
  \cellcolor{green!20}\textbf{1} &
  \cellcolor{green!20}\textbf{5.74 (6.84)} &
  \cellcolor{green!20}2 &
  \cellcolor{green!20}1.61 (0.31) &
  \cellcolor{green!20}\textbf{1} &
  \cellcolor{green!20}\textbf{1.66 (0.57)} &
  \cellcolor{green!20}\textbf{1} &
  \cellcolor{green!20}\textbf{1.56 (0.20)} &
  \cellcolor{green!20}\textbf{1} &
  \cellcolor{green!20}\textbf{4.39 (0.41)} &
  \cellcolor{green!20}\textbf{1} &
  \cellcolor{green!20}\textbf{1.12 (0.08)} &
  \cellcolor{green!20}\textbf{1} &
  \cellcolor{green!20}\textbf{11.83 (4.53)} \\ \hline
 &
  \texttt{DeepPerf} &
  \textbf{1} &
  \textbf{6.85 (1.60)} &
  2 &
  9.06 (9.76) &
  3 &
  1.70 (0.37) &
  2 &
  1.74 (0.98) &
  2 &
  2.23 (0.39) &
  2 &
  3.58 (1.00) &
  2 &
  1.54 (0.40) &
  2 &
  15.55 (1.71) \\
 &
  \texttt{DECART} &
  2 &
  7.47 (0.72) &
  \textbf{1} &
  \textbf{5.17 (5.07)} &
  \cellcolor{red!20}\textbf{1} &
  \cellcolor{red!20}\textbf{1.47 (0.12)} &
  3 &
  3.66 (1.42) &
  N/A &
  N/A &
  N/A &
  N/A &
  N/A &
  N/A &
  N/A &
  N/A \\
 &
  \texttt{Perf-AL} &
  4 &
  30.67 (6.84) &
  3 &
  69.50 (2.49) &
  5 &
  37.74 (2.59) &
  5 &
  36.41 (6.47) &
  4 &
  66.63 (0.14) &
  4 &
  31.98 (0.06) &
  4 &
  59.99 (0.27) &
  3 &
  58.48 (0.21) \\
 &
  \texttt{SPLConqueror} &
  3 &
  13.84 (0.00) &
  4 &
  231.68 (0.00) &
  4 &
  29.01 (0.00) &
  4 &
  6.32 (0.00) &
  3 &
  33.73 (0.00) &
  3 &
  17.42 (0.00) &
  3 &
  37.14 (0.00) &
  4 &
  236.48 (0.00) \\
\multirow{-5}{*}{Size 4} &
  \Model &
  \cellcolor{green!20}\textbf{1} &
  \cellcolor{green!20}\textbf{6.59 (1.77)} &
  \cellcolor{green!20}\textbf{1} &
  \cellcolor{green!20}\textbf{3.68 (2.78)} &
  2 &
  1.61 (0.29) &
  \cellcolor{green!20}\textbf{1} &
  \cellcolor{green!20}\textbf{1.13 (0.63)} &
  \cellcolor{green!20}\textbf{1} &
  \cellcolor{green!20}\textbf{1.49 (0.15)} &
  \cellcolor{green!20}\textbf{1} &
  \cellcolor{green!20}\textbf{3.22 (0.35)} &
  \cellcolor{green!20}\textbf{1} &
  \cellcolor{green!20}\textbf{1.10 (0.06)} &
  \cellcolor{green!20}\textbf{1} &
  \cellcolor{green!20}\textbf{10.10 (3.23)} \\ \hline
 &
  \texttt{DeepPerf} &
  2 &
  6.66 (1.61) &
  2 &
  7.08 (8.03) &
  2 &
  1.69 (0.48) &
  2 &
  1.44 (0.90) &
  2 &
  1.94 (0.68) &
  2 &
  2.82 (0.57) &
  2 &
  1.45 (0.23) &
  2 &
  10.17 (0.91) \\
 &
  \texttt{DECART} &
  3 &
  7.14 (0.89) &
  \textbf{1} &
  \textbf{3.54 (1.92)} &
  \cellcolor{red!20}\textbf{1} &
  \cellcolor{red!20}\textbf{1.37 (0.26)} &
  3 &
  2.15 (1.05) &
  N/A &
  N/A &
  N/A &
  N/A &
  N/A &
  N/A &
  N/A &
  N/A \\
 &
  \texttt{Perf-AL} &
  5 &
  30.45 (4.91) &
  3 &
  69.29 (0.69) &
  4 &
  35.76 (4.47) &
  5 &
  36.20 (3.75) &
  4 &
  66.59 (0.20) &
  4 &
  31.97 (0.14) &
  4 &
  59.95 (0.68) &
  3 &
  58.39 (0.21) \\
 &
  \texttt{SPLConqueror} &
  4 &
  14.06 (0.00) &
  4 &
  296.68 (0.00) &
  3 &
  29.87 (0.00) &
  4 &
  6.36 (0.00) &
  3 &
  35.86 (0.00) &
  3 &
  17.69 (0.00) &
  3 &
  36.21 (0.00) &
  4 &
  208.08 (0.00) \\ 
\multirow{-5}{*}{Size 5} &
  \Model &
  \cellcolor{green!20}\textbf{1} &
  \cellcolor{green!20}\textbf{5.96 (2.07)} &
  \cellcolor{green!20}\textbf{1} &
  \cellcolor{green!20}\textbf{2.15 (1.52)} &
  \textbf{1} &
  \textbf{1.46 (0.26)} &
  \cellcolor{green!20}\textbf{1} &
  \cellcolor{green!20}\textbf{0.79 (0.33)} &
  \cellcolor{green!20}\textbf{1} &
  \cellcolor{green!20}\textbf{1.16 (0.08)} &
  \cellcolor{green!20}\textbf{1} &
  \cellcolor{green!20}\textbf{2.39 (0.22)} &
  \cellcolor{green!20}\textbf{1} &
  \cellcolor{green!20}\textbf{1.07 (0.05)} &
  \cellcolor{green!20}\textbf{1} &
  \cellcolor{green!20}\textbf{6.60 (1.34)} \\
\bottomrule
\end{tabular}
}

%% file: Tables/compare_models.tex
\begin{tabular}{P{1cm}P{1cm}P{1cm}P{1cm}|P{1cm}P{1cm}|P{1cm}P{1cm}|P{1cm}P{1cm}|P{1cm}P{1cm}|P{1cm}P{1cm}|P{1cm}P{1cm}}
\toprule

\textbf{System} &
   &
  \textbf{\Model} &
  \textbf{rDNN} &
  \textbf{\Model$_{RF}$} &
  \textbf{RF} &
  \textbf{\Model$_{CART}$} &
  \textbf{{CART}} &
  \textbf{\Model$_{LR}$} &
  \textbf{LR} &
  \textbf{\Model$_{SVR}$} &
  \textbf{SVR} &
  \textbf{\Model$_{KRR}$} &
  \textbf{KRR} &
  \textbf{\Model$_{kNN}$} &
  \textbf{$k$NN} \\ \midrule
\multirow{5}{*}{\textsc{Apache}}      & Size 1 & \cellcolor{green!20}\textbf{1} & 2 & \cellcolor{green!20}\textbf{1}& 3 & 2 & 2 & \cellcolor{green!20}\textbf{1} & 5  & 4 & 4  & 6  & 7  & 2  & 2  \\
                                      & Size 2 & 2          & \cellcolor{green!20}\textbf{1}& 2 & 2 & 2 & \cellcolor{green!20}\textbf{1}& 3  & 5  & 6 & 7  & 8  & 9  & 4  & 6  \\
                                      & Size 3 & \cellcolor{green!20}\textbf{1} & 2 & 3 & 2 & 4 & 4 & \cellcolor{green!20}\textbf{1} & 7  & 6 & 9  & 10 & 11 & 5  & 8  \\
                                      & Size 4 & 2          & 2 & 3 & 2 & 5 & 4 & \cellcolor{green!20}\textbf{1} & 8  & 7 & 9  & 10 & 11 & 6  & 8  \\
                                      & Size 5 & 2          & 4 & 4 & 3 & 5 & 5 & \cellcolor{green!20}\textbf{1} & 9  & 7 & 10 & 11 & 12 & 6  & 8  \\ \hline
\multirow{5}{*}{\textsc{BDB-C}}       & Size 1 & 2          & 2 & 2 & 6 & \cellcolor{green!20}\textbf{1}& \cellcolor{green!20}\textbf{1}& 3  & 8  & 3 & 4  & 2  & 7  & 3  & 5  \\
                                      & Size 2 & \cellcolor{green!20}\textbf{1} & \cellcolor{green!20}\textbf{1}& 3 & 4 & \cellcolor{green!20}\textbf{1}& 2 & 7  & 6  & 5 & 5  & 4  & 6  & 5  & 6  \\
                                      & Size 3 & 2          & 5 & 4 & 4 & 3 & \cellcolor{green!20}\textbf{1}& 6  & 9  & 8 & 7  & 6  & 9  & 8  & 9  \\
                                      & Size 4 & \cellcolor{green!20}\textbf{1} & 3 & 3 & 4 & \cellcolor{green!20}\textbf{1}& 2 & 5  & 9  & 6 & 6  & 5  & 8  & 6  & 7  \\
                                      & Size 5 & \cellcolor{green!20}\textbf{1} & 3 & 2 & 4 & \cellcolor{green!20}\textbf{1}& \cellcolor{green!20}\textbf{1}& 4  & 9  & 6 & 7  & 5  & 9  & 6  & 8  \\ \hline
\multirow{5}{*}{\textsc{BDB-J}}       & Size 1 & 2          & 2 & 2 & 3 & \cellcolor{green!20}\textbf{1}& \cellcolor{green!20}\textbf{1}& 9  & 8  & 6 & 4  & 3  & 7  & 5  & 5  \\
                                      & Size 2 & 4          & 3 & \cellcolor{green!20}\textbf{1}& \cellcolor{green!20}\textbf{1}& 3 & 2 & 4  & 10 & 8 & 5  & 4  & 9  & 6  & 7  \\
                                      & Size 3 & 2          & 3 & \cellcolor{green!20}\textbf{1}& \cellcolor{green!20}\textbf{1}& 2 & 2 & 3  & 8  & 6 & 5  & 3  & 7  & 4  & 4  \\
                                      & Size 4 & 2          & 4 & \cellcolor{green!20}\textbf{1}& \cellcolor{green!20}\textbf{1}& 3 & 3 & 4  & 9  & 8 & 8  & 5  & 9  & 7  & 6  \\
                                      & Size 5 & 2          & 4 & \cellcolor{green!20}\textbf{1}& \cellcolor{green!20}\textbf{1}& 3 & 3 & 4  & 9  & 7 & 8  & 5  & 9  & 6  & 6  \\ \hline
\multirow{5}{*}{\textsc{x264}}        & Size 1 & \cellcolor{green!20}\textbf{1} & \cellcolor{green!20}\textbf{1}& 3 & 2 & 3 & 3 & 4  & 5  & 5 & 9  & 8  & 6  & 4  & 7  \\
                                      & Size 2 & \cellcolor{green!20}\textbf{1} & \cellcolor{green!20}\textbf{1}& 3 & 3 & 3 & 3 & 2  & 4  & 7 & 9  & 4  & 5  & 6  & 8  \\
                                      & Size 3 & \cellcolor{green!20}\textbf{1} & 2 & 5 & 5 & 4 & 4 & 2  & 5  & 6 & 9  & 5  & 7  & 8  & 9  \\
                                      & Size 4 & \cellcolor{green!20}\textbf{1} & 2 & 5 & 6 & 3 & 4 & 3  & 7  & 7 & 10 & 6  & 8  & 9  & 11 \\
                                      & Size 5 & \cellcolor{green!20}\textbf{1} & 2 & 5 & 6 & 3 & 4 & 6  & 8  & 8 & 11 & 7  & 9  & 10 & 12 \\ \hline
\multirow{5}{*}{\textsc{HSMGP}}       & Size 1 & \cellcolor{green!20}\textbf{1} & 8 & 6 & 5 & 7 & 7 & 2  & 11 & 4 & 5  & 3  & 10 & 8  & 9  \\
                                      & Size 2 & \cellcolor{green!20}\textbf{1} & 2 & 5 & 6 & 8 & 8 & 2  & 10 & 7 & 8  & 4  & 10 & 9  & 9  \\
                                      & Size 3 & \cellcolor{green!20}\textbf{1} & 2 & 4 & 5 & 7 & 7 & 2  & 10 & 6 & 9  & 3  & 10 & 8  & 8  \\
                                      & Size 4 & \cellcolor{green!20}\textbf{1} & 2 & 3 & 4 & 7 & 8 & 4  & 11 & 6 & 10 & 5  & 11 & 9  & 9  \\
                                      & Size 5 & \cellcolor{green!20}\textbf{1} & 2 & 3 & 3 & 7 & 6 & 10 & 9  & 4 & 8  & 5  & 10 & 9  & 9  \\ \hline
\multirow{5}{*}{\textsc{HIPA$^{cc}$}} & Size 1 & \cellcolor{green!20}\textbf{1} & 2 & 3 & 3 & 4 & 4 & 9  & 8  & 5 & 7  & 4  & 9  & 6  & 8  \\
                                      & Size 2 & \cellcolor{green!20}\textbf{1} & 2 & 3 & 3 & 3 & 3 & 6  & 10 & 5 & 6  & 5  & 8  & 4  & 7  \\
                                      & Size 3 & \cellcolor{green!20}\textbf{1} & 2 & 3 & 3 & 3 & 3 & 5  & 11 & 5 & 8  & 6  & 10 & 4  & 9  \\
                                      & Size 4 & \cellcolor{green!20}\textbf{1} & 2 & 4 & 4 & 3 & 3 & 8  & 11 & 7 & 10 & 8  & 11 & 6  & 5  \\
                                      & Size 5 & \cellcolor{green!20}\textbf{1} & 2 & 3 & 2 & 2 & 2 & 10 & 12 & 7 & 8  & 9  & 11 & 5  & 6  \\ \hline
\multirow{5}{*}{\textsc{VP8}}         & Size 1 & \cellcolor{green!20}\textbf{1} & 5 & 4 & 6 & 2 & 3 & 5  & 7  & 9 & 10 & 7  & 12 & 10 & 11 \\
                                      & Size 2 & 4          & 5 & 3 & 4 & \cellcolor{green!20}\textbf{1}& 2 & 4  & 8  & 6 & 11 & 7  & 12 & 9  & 10 \\
                                      & Size 3 & \cellcolor{green!20}\textbf{1} & 2 & 2 & 2 & 2 & 3 & 5  & 11 & 5 & 9  & 6  & 10 & 8  & 8  \\
                                      & Size 4 & \cellcolor{green!20}\textbf{1} & 2 & 2 & 2 & 2 & 2 & 3  & 8  & 4 & 6  & 3  & 7  & 5  & 6  \\
                                      & Size 5 & \cellcolor{green!20}\textbf{1} & 2 & 3 & 4 & 2 & 3 & 6  & 12 & 7 & 10 & 5  & 11 & 8  & 9  \\ \hline
\multirow{5}{*}{\textsc{Lrzip}}       & Size 1 & 3          & 4 & 2 & 3 & \cellcolor{green!20}\textbf{1}& 2 & 5  & 10 & 5 & 5  & 7  & 9  & 6  & 5  \\
                                      & Size 2 & 3          & 4 & 2 & 2 & \cellcolor{green!20}\textbf{1}& 2 & 5  & 8  & 6 & 6  & 7  & 9  & 5  & 5  \\
                                      & Size 3 & 3          & 4 & 2 & \cellcolor{green!20}\textbf{1}& \cellcolor{green!20}\textbf{1}& 2 & 6  & 9  & 7 & 6  & 8  & 10 & 5  & 5  \\
                                      & Size 4 & 3          & 4 & 4 & 7 & \cellcolor{green!20}\textbf{1}& 2 & 8  & 10 & 2 & 3  & 6  & 9  & 5  & 8  \\
                                      & Size 5 & 3          & 4 & \cellcolor{green!20}\textbf{1}& 3 & \cellcolor{green!20}\textbf{1}& 2 & 7  & 9  & 7 & 8  & 9  & 11 & 5  & 6  \\ \midrule
Average &
   &
  \cellcolor{green!20}\textbf{1.63} &
  2.78 &
  2.90 &
  3.38 &
  2.95 &
  3.15 &
  4.63 &
  8.58 &
  6.00 &
  7.48 &
  5.85 &
  9.13 &
  6.25 &
  7.35
\\ \bottomrule

\end{tabular}

%% file: Figures/Sensitivity_to_depth.tex
\hspace{0.1cm}
\begin{subfigure}{.42\columnwidth}
  \centering
  
\includegraphics[width=\linewidth]{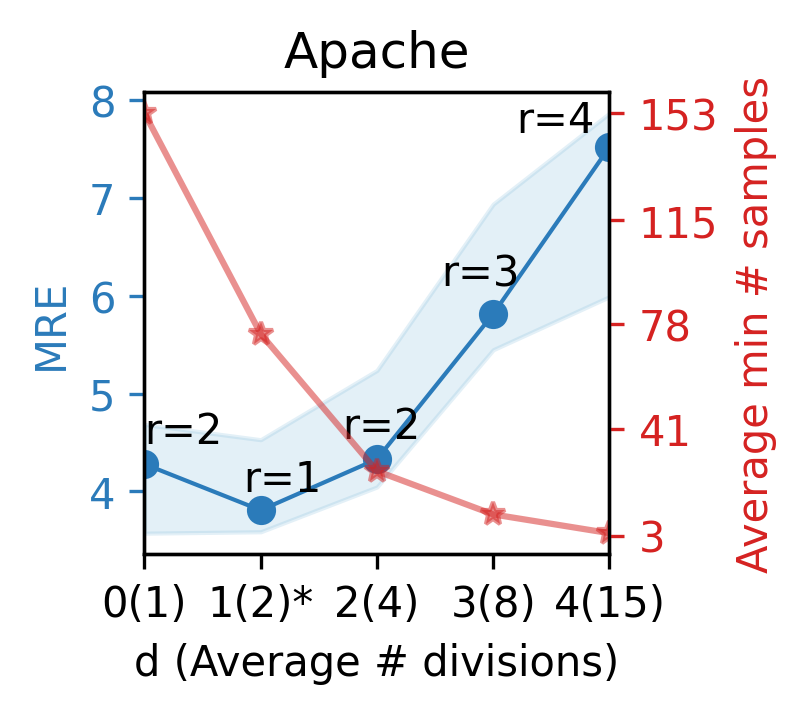}  
  \label{fig:depth-Apache}
\end{subfigure}
~\hspace{0.1cm}
\begin{subfigure}{.43\columnwidth}
  \centering
  \includegraphics[width=\linewidth]{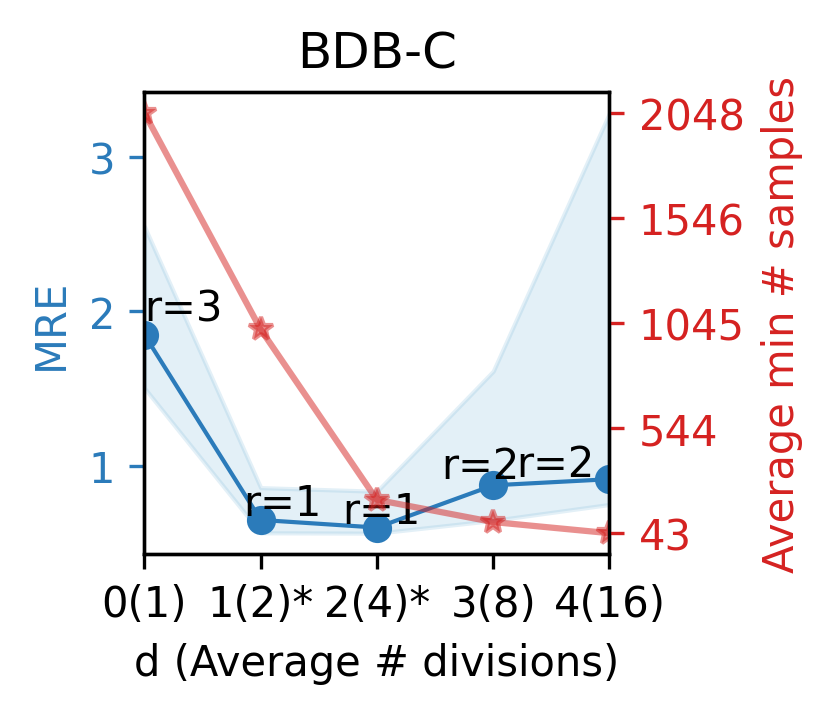} 
  \label{fig:depth-BDBC}
\end{subfigure}
~\hspace{0.1cm}
\begin{subfigure}{.42\columnwidth}
  \centering
  \includegraphics[width=\linewidth]{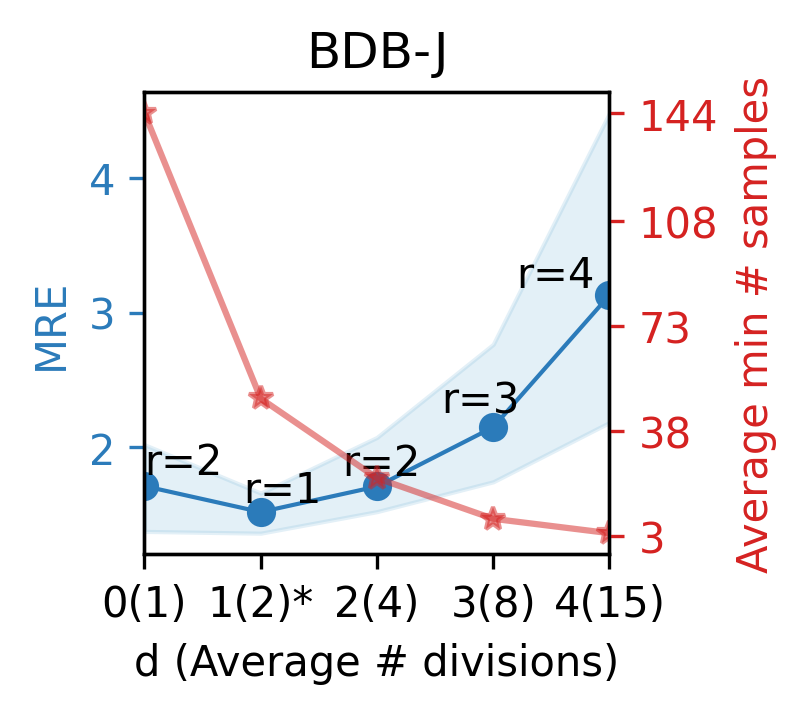} 
  \label{fig:depth-BDBJ}
\end{subfigure}
~
\begin{subfigure}{.43\columnwidth}
  \centering
  \includegraphics[width=\linewidth]{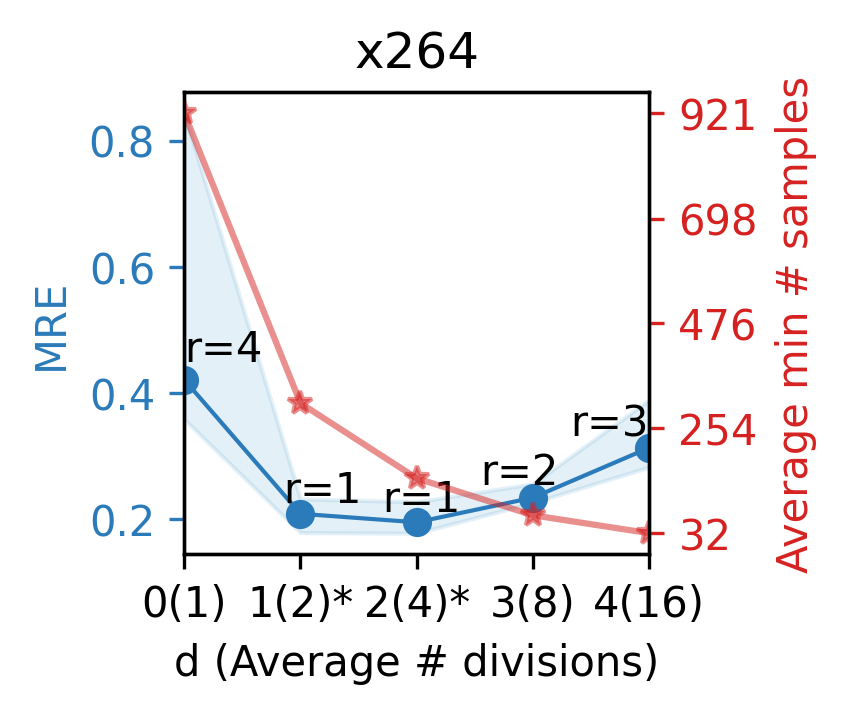}  
  \label{fig:depth-x264}
\end{subfigure}

\vspace{-0.3cm}

\begin{subfigure}{.45\columnwidth}
  \centering
  \includegraphics[width=\linewidth]{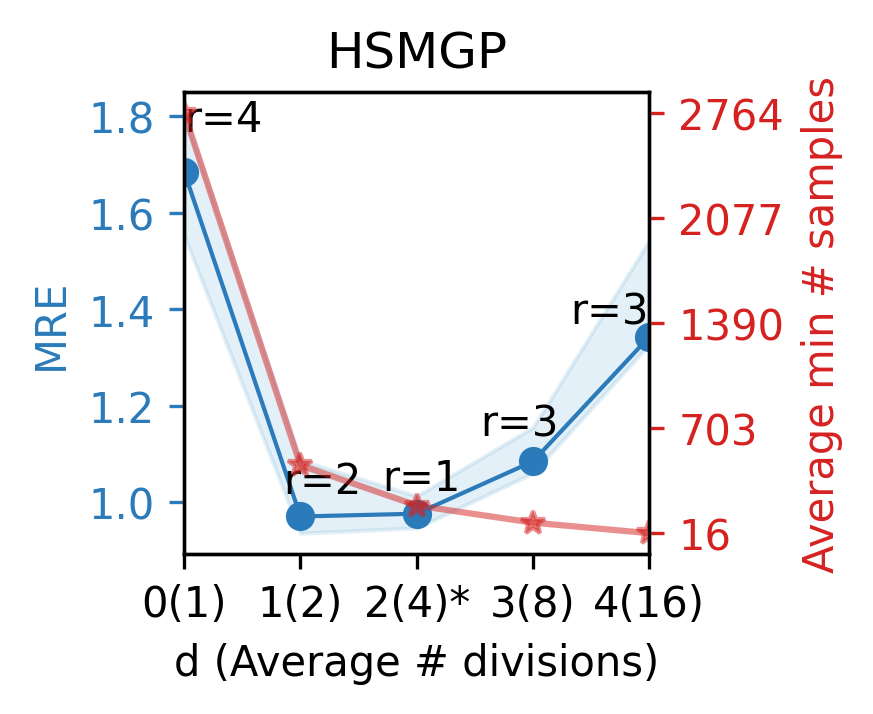} 
  \label{fig:depth-Hsmgp}
\end{subfigure}
~
\begin{subfigure}{.46\columnwidth}
  \centering
  \includegraphics[width=\linewidth]{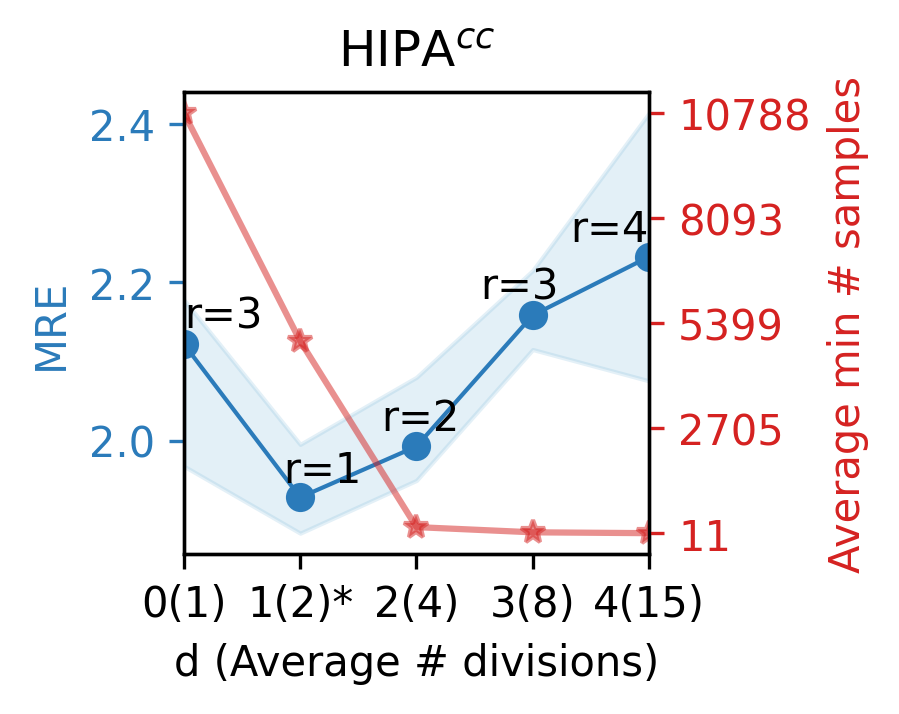}  
  \label{fig:depth-hipacc}
\end{subfigure}
~
\begin{subfigure}{.46\columnwidth}
  \centering
  \includegraphics[width=\linewidth]{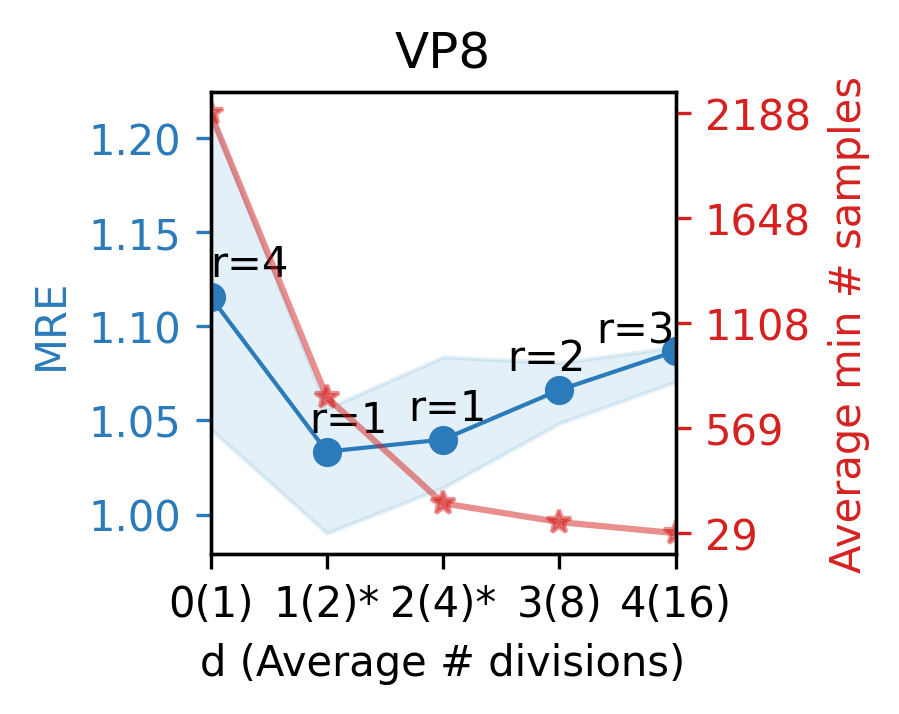} 
  \label{fig:depth-VP8}
\end{subfigure}
~
\begin{subfigure}{.42\columnwidth}
  \centering
  \includegraphics[width=\linewidth]{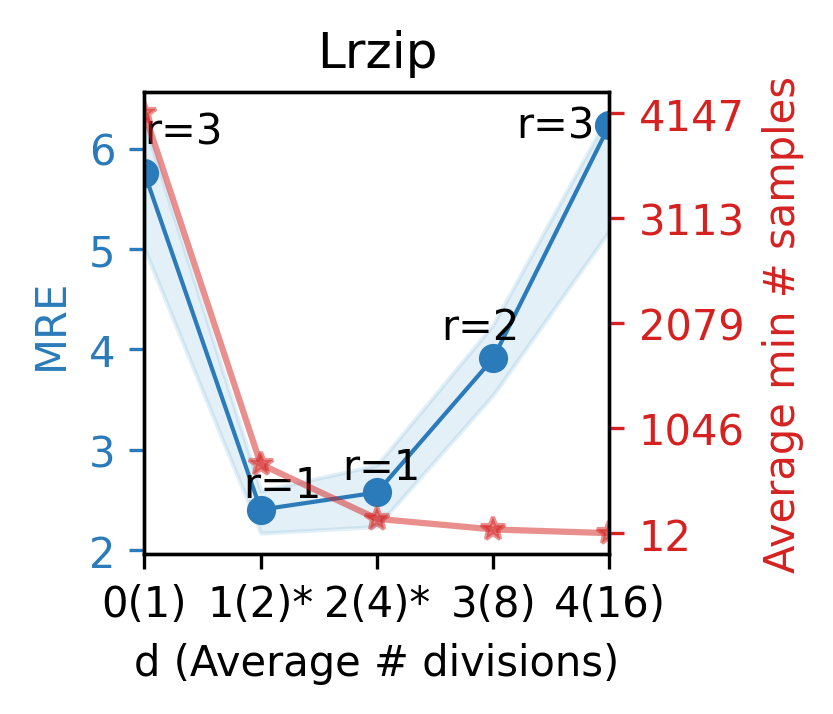} 
  \label{fig:depth-Lrzip}
\end{subfigure}

%% file: Tables/min_max_time.tex
\begin{table}[t!]
\caption{The overhead ranges across all systems and sizes.}
\centering
\footnotesize
\begin{adjustbox}{width=\columnwidth,center}
\begin{tabular}{lll}
\toprule
\textbf{Approach}             & \textbf{Overhead (min)}                  & \textbf{Restriction and Prerequisite}               \\
\midrule
\texttt{DeepPerf} & 3 to 60 & None \\
\texttt{DECART} & 0.07 to 0.5 & does not work on mixed systems \\
\texttt{Perf-AL} & 0.08 to 1 & None \\
\texttt{SPLConqueror} & 4$\times 10^{-4}$ to 5$\times 10^{-3}$ & needs to select sampling method(s) \\
\Model & 6 to 56 & needs to set the depth $d$ \\
--- \Model~(\textit{dividing}) & 9$\times 10^{-4}$ to 0.18 & None \\
--- \Model~(\textit{training}) & 4 to 52 & None \\
--- \Model~(\textit{predicting}) & 1.3 to 5 & None \\

\bottomrule
\end{tabular}
\end{adjustbox}
\label{tb:max_min}
\end{table}

%% file: discussion.tex
\section{Discussion}
\label{sec:discussion}

\subsection{Why does \Model~Work?}

\begin{figure}[!t]
\centering
\footnotesize

\begin{subfigure}{.48\columnwidth}
  \centering
  \includegraphics[width=\linewidth]{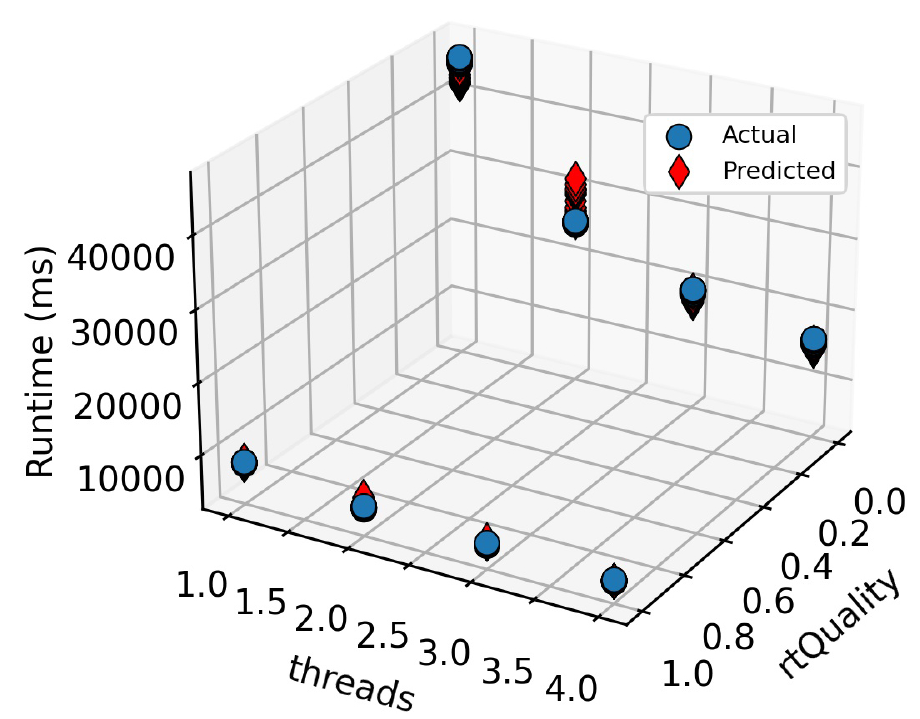} 
  \caption{prediction by \Model}
\end{subfigure}
\begin{subfigure}{.50\columnwidth}
  \centering
  \includegraphics[width=\linewidth]{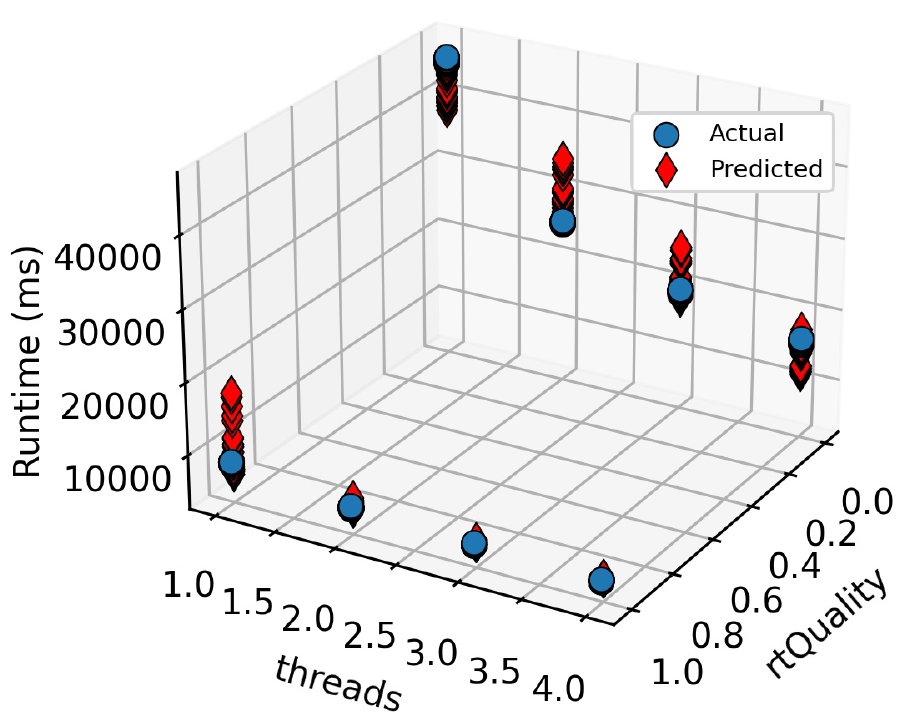} 
  \caption{prediction by \texttt{DeepPerf}}
\end{subfigure}

\caption{Example run of the actual and predicted performance by \Model~and \texttt{DeepPerf} for \textsc{VP8}.}
\label{fig:discussion_scatter}
\end{figure}

To provide a more detailed understanding of why \Model~performs better than state-of-the-art, in Figure~\ref{fig:discussion_scatter}, we showcase the most common run of the predicted performance by \Model~and \texttt{DeepPerf} against actual performance. Clearly, we note that the sample sparsity is rather obvious where there are two distant divisions. \texttt{DeepPerf}, as an approach that relies on a single and global rDNN, has been severely affected by such highly sparse samples: we see that the model tries to cover points in both divisions, but fails to do so as it tends to overfit the points in one or the other. This is why, in Figure~\ref{fig:discussion_scatter}b, its prediction on some configurations that should lead to low runtime tend to have much higher values (e.g., when \texttt{rtQuality=1} and \texttt{threads=1}) while some of those that should have high runtime may be predicted with much lower values (e.g., when \texttt{rtQuality=0} and \texttt{threads=1}). \Model, in contrast, handles such a sample sparsity well as it contains different local models that particularly cater to each division identified, hence leading to high accuracy (Figure~\ref{fig:discussion_scatter}a).

\subsection{Strengths and Limitations}

The first strength of \Model~is that the concept of ``divide-and-learn'', paired with the rDNN, can handle both sample sparsity and feature sparsity well. As from Section~\ref{subsec:rq1} for \textbf{RQ1}, this has led to better accuracy and better utilization of the sample data than the state-of-the-art approaches.


The second strength is that, as from Section~\ref{subsec:rq2} for \textbf{RQ2}, \Model~can improve different local models compared with when they are used alone as a global model. While we set rDNN as the default for the best accuracy, one can also easily replace it with others such as LR for faster training and better interoperability. This enables great flexibility with \Model~to make trade-offs on different concerns of the practical scenarios.


A limitation of \Model~is that it takes a longer time to build the model than some state-of-the-art approaches. On a machine with CPU 2GHz and 16GB RAM, \Model~needs between 6 and 56 minutes for systems with up to 33 options and more than 2,000 samples.


\subsection{Why $d \in \{1|2\}$ is Highly Effective?}

We have shown that the setting of $d$ in \Model~should be neither too small nor too large; the key intention behind the $d$ is to reach a good balance between handling the sample sparsity and providing sufficient data for the local models to generalize. This is especially true when the CART might produce divisions with imbalanced sample sizes, e.g., we observed cases where there is a division with around 500 samples while one other has merely less than 10. Our experimental results show that such ``sweet points'' tend to be $d=1$ or $d=2$ for the cases studied in this work.

However, the notion of ``too small'' and ``too large'' should be interpreted cautiously depending on the systems and data size. That is, although in this study, setting $d=1$ or $d=2$ appears to be appropriate; they might become ``too small'' settings when the data size increases considerably and/or the system naturally exhibits well-balanced divisions of configuration samples in the landscapes. Yet, the pattern of quadratic correlation between $d$ and the error of \Model~should remain unchanged.


\subsection{Using \Model~in Practice}

Like many other data-driven approaches, using \Model~is straightforward and free of assumptions about the software systems, data, and environments. We would recommend setting $d=1$ or $d=2$ by default, especially when the data sample size is similar to those we studied in this work. Of course, it is always possible to fine-tune the $d$ value by training \Model~with alternative settings under the configuration samples available. Given the quadratic correlation between $d$ and the error, it is possible to design a simple heuristic for this, e.g., we compare the accuracy of \Model~trained with $d=i$ and $d=i+1$ starting from $d=1$ and finally selecting the maximum $d$ value $k$ such that \Model~with $k+1$ is less accurate than \Model~with $k$.

\subsection{Threats to Validity}

{\textbf{Internal Threats.}} Internal threats to validity are related to the parameters used. In this work, we set the same setting as used in state-of-the-art studies~\cite{DBLP:journals/sqj/SiegmundRKKAS12,DBLP:conf/kbse/GuoCASW13,DBLP:conf/icse/HaZ19,DBLP:conf/esem/ShuS0X20}. We have also shown the sensitivity of \Model~to $d$ and reveal that there exists a generally best setting. We repeat the experiments for 30 runs and use Scott-Knott test for multiple comparisons.

{\textbf{Construct Threats.}} Threats to construct validity may lie in the metric used. In this study, MRE is chosen for two reasons: (1) it is a relative metric and hence is insensitive to the scale of the performance; (2) MRE has been recommended for performance prediction by many latest studies~\cite{DBLP:conf/icse/HaZ19, DBLP:conf/esem/ShuS0X20, DBLP:journals/ese/GuoYSASVCWY18}.

{\textbf{External Threats.}} External validity could be raised from the subject systems and training samples used. To mitigate such, we evaluate eight commonly used subject systems selected from the latest studies. We have also examined different training sample sizes as determined by \texttt{SPLConqueror}~\cite{DBLP:journals/sqj/SiegmundRKKAS12}---a typical method. Yet, we agree that using more subject systems and data sizes may be fruitful, especially for examining the sensitivity of $d$ which may lead to a different conclusion when there is a much larger set of training configuration samples than we consider in this study.

%% file: related.tex
\section{Related Work}
\label{sec:related}

We now discuss the related work in light of \Model.

\textbf{Analytical model.} Predicting software performance can be done by analyzing the code structure and architecture of the systems~\cite{di2004compositional,DBLP:conf/icse/VelezJSAK21}. For example, Marco and Inverardi~\cite{di2004compositional} apply queuing network to model the latency of requests processed by the software. Velez \textit{et al.}~\cite{DBLP:conf/icse/VelezJSAK21} use local measurements and
dynamic taint analysis to build a model that can predict performance for part of the code. However, analytical models require full understanding and access to the software's internal states, which may not always be possible/feasible. \Model~is not limited to those scenarios as it is a data-driven approach.


\textbf{Statistical learning-based model.} Data-driven learning has relied on various statistical models, such as linear regressions~\cite{DBLP:journals/sqj/SiegmundRKKAS12,DBLP:conf/sigsoft/SiegmundGAK15,DBLP:journals/tc/SunSZZC20,DBLP:journals/is/KangKSGL20}, tree-liked model~\cite{DBLP:conf/icdcs/HsuNFM18,DBLP:conf/kbse/SarkarGSAC15,DBLP:journals/corr/abs-1801-02175}, and fourier-learning models~\cite{zhang2015performance,DBLP:conf/icsm/Ha019}, \textit{etc}. Among others, \texttt{SPLConqueror}~\cite{DBLP:journals/sqj/SiegmundRKKAS12} utilizes linear regression
combined with different sampling methods and a step-wise feature selection to capture the interactions between configuration options. \texttt{DECART}~\cite{DBLP:journals/ese/GuoYSASVCWY18} is an improved CART with an efficient sampling method~\cite{zhang2015performance}. However, recent work reveals that those approaches do not work well with small datasets~\cite{DBLP:conf/icse/HaZ19}, which is rather common for configurable software systems due to their expensive measurements. This is a consequence of not fully handling the sparsity in configuration data. Further, they come with various restrictions, e.g., \texttt{DECART} does not work on mixed systems while \texttt{SPLConqueror} needs an extensive selection of the right sampling method(s). In contrast, we showed that \Model~produces significantly more accurate results while does not limit to those restrictions.

\textbf{Ensemble model.} Models can be combined in a shared manner to predict software performance. For example, Chen and Bahsoon~\cite{DBLP:journals/tse/ChenB17} propose an ensemble approach, paired with feature selection for mitigating feature sparsity, to model software performance. Other classic ensemble learning models such as Bagging~\cite{breiman1996bagging} and Boosting~\cite{schapire2003boosting} (e.g., RF) can also be equally adopted. Indeed, at a glance, our \Model~does seem similar to the ensemble model as they all maintain a pool of local models. However, the key difference is that the classic ensemble models will inevitably share information between the local models at one or more of the following levels:
\begin{itemize}
    \item At the training level, e.g., the local models in Boosting learn the same samples but with a different focus; the Bucket of Models (i.e., what Chen and Bahsoon~\cite{DBLP:journals/tse/ChenB17} did) builds local models on the same data and uses the best upon prediction.
    \item At the model prediction level, e.g., Bagging aggregates the results of local models upon prediction.
\end{itemize}

\Model, in contrast, has no information sharing throughout the learning as the samples are split and so does the prediction of the local models. This has enabled it to better isolate the samples and cope with their inherited sparsity, e.g., recall from \textbf{RQ2}, the overall accuracy of \texttt{DaL$_{CART}$} is better than RF (they both use CART as the local models but learn with and without sharing information).


\textbf{Deep learning-based model.} A variety of studies apply neural network with multiple layers and/or ensemble learning to predict software performance~\cite{DBLP:conf/icse/HaZ19, DBLP:conf/esem/ShuS0X20,DBLP:conf/sbac-pad/NemirovskyAMNUC17,app11083706,DBLP:journals/concurrency/FalchE17,DBLP:conf/iccad/KimMMSR17,DBLP:conf/sc/MaratheAJBTKYRG17,DBLP:conf/im/JohnssonMS19,DBLP:journals/jsa/ZhangLWWZH18}. \texttt{DeepPerf}~\cite{DBLP:conf/icse/HaZ19} is a state-of-the-art DNN model with $L_{1}$ regularization to mitigate feature sparsity for any configurable systems, and it can be more accurate than many other existing approaches. The most recently proposed \texttt{Perf-AL}~\cite{DBLP:conf/esem/ShuS0X20} relied on adversarial learning, which consists of a generative network to predict the performance and a discriminator network to distinguish the predictions and the actual labels. Nevertheless, existing deep learning approaches capture only the feature sparsity while ignoring the sample sparsity, causing serve risks of overfitting even with regularization in place. Compared with those, we have shown that, by capturing sample sparsity, \Model~is able to improve the accuracy considerably with better efficiency and acceptable overhead.



\textbf{Hybrid model.} 
The analytical models can be combined with data-driven ones to form a hybrid model~\cite{DBLP:conf/wosp/HanYP21,didona2015enhancing,DBLP:conf/icse/WeberAS21}. Among others, Didona \textit{et al.}~\cite{didona2015enhancing} use linear regression and $k$NN to learn certain components of a queuing network. Conversely, Weber \textit{et al.}~\cite{DBLP:conf/icse/WeberAS21} propose to learn the performance of systems based on the parsed source codes from the system to the function level. We see \Model~as being complementary to those hybrid models due to its flexibility in selecting the local model: when needed, the local models can be replaced with hybrid ones, making itself a hybrid variant. In case the internal structure of the system is unknown, \Model~can also work in its default as a purely data-driven approach.

%% file: conclusion.tex
\section{Conclusion}
\label{sec:conclusion}

This paper proposes \Model, an approach that effectively handles the sparsity issues in configurable software performance prediction. By formulating a classification problem with pseudo labels on top of the original regression problem, \Model~extracts the branches/leaves from a CART which divides the samples of configuration into distant divisions and trains a dedicated local rDNN for each division thereafter. Prediction of the new configuration is then made by the rDNN of division inferred by a Random Forest classifier. As such, the division of samples and the trained local model handles the sample sparsity while the rDNN deals with the feature sparsity.

We evaluate \Model~on eight real-world systems that are of diverse domains and scales, together with five sets of training data. The results show that \Model~is:

\begin{itemize}
    \item \textbf{effective} as it is competitive to the best state-of-the-art approach on 33 out of 40 cases, in which 26 of them are significantly better with up to $1.94\times$ MRE improvement; 
    \item \textbf{efficient} since it often requires fewer samples to reach the same/better accuracy compared with the state-of-the-art; 
    \item \textbf{flexible} given that it considerably improves various global models when they are used as the local model therein; 
    \item \textbf{robust} because, given the quadratic correlation, a middle $d$ value(s) (between 0 and the bound set by CART) can be robust and leads to the best accuracy across the cases, e.g., $d=1$ or $d=2$ under the sample sizes in this work.
\end{itemize}


Mitigating the issues caused by sparsity is only one step towards better performance prediction, hence the possible future work based on \Model~is vast, including multi-task prediction of performance under different environments and merging diverse local models (e.g., a mix of rDNN and LR) as part of the ``divide-and-learn'' concept. Consolidating \Model~with an adaptive $d$ is also within our agenda.

\section{Data Availability}
\label{sec:data}

Data, code, and supplementary figures of this work can be found at our repository: \texttt{\textcolor{blue}{\url{https://github.com/ideas-labo/DaL}}}.